**Electron-Lattice Interaction and its Impact on High $T_c$**

**Superconductivity**

V. Z.  Kresin
Lawrence Berkeley Laboratory, University of California, Berkeley,
    California, 94720, USA

S. A. Wolf

 Department of Materials Science and Engineering and Department of
    Physics,University of Virginia, Charlottesville,Virginia 22904,USA

Abstract

In this Colloquium , the main features of the electron-lattice interaction are discussed and high values of  the critical temperature up to room temperature could be provided. While the issue of the mechanism of superconductivity in the high $T_c$ cuprates continues to be controversial, one can state that there have been many experimental results demonstrating that the lattice makes a strong impact on the pairing of electrons. The polaronic nature of the carriers is also a manifestation of strong electron-lattice interaction. One can propose an experiment that allows an unambiguous determination of the intermediate boson (phonon, magnon, exciton, etc.) which provides the pairing. The electron-lattice interaction increases for nanosystems, and this is due to an effective increase in the density of states.

**Outline.**

I. Introduction

II. Electron-lattice interaction and upper limit of $T_c$

III. Experimental data and analysis

    A. Tunneling spectroscopy

        1. McMillan-Rowell method

        2. Tunneling studies of the cuprates

    B. Infrared spectroscopy

    C. Photoemission and ultrafast electron spectroscopy

    D. Isotope effect

    E. Heat capacity

IV. Polaronic effects

    A. Polarons and isotope effects

    B. "Local" pairs: bipolarons, U-centers, and the BEC-BCS scheme

V. Phonon-plasmon mechanism

VI. Electron-phonon interaction and the "pseudogap" state

VII. Proposed experiment

VIII. Superconducting state of nanoclusters

    A. Nanoparticles: size quantization.

    B. Nanoclusters and the high $T_c$ state

IX. Conclusion

Acknowledgements

References

I. Introduction

This Colloquium addresses the current experimental and theoretical situation concerning the importance of the interaction between electrons and the crystal lattice in novel superconducting systems, especially in high $T_c$ cuprates. It will be demonstrated that the electron-lattice interaction is an important factor underlying the nature of high $T_c$ superconductivity..

The phenomenon of superconductivity was discovered by Kamerlingh-Onnes in 1911 (Onnes,1911), and presently we are approaching the 100th anniversary of this event . The phenomenon was explained only in 1957 by Bardeen, Cooper, and Schrieffer (BCS). According to the classical BCS theory, the key phenomenon occurring in superconductors is the pairing of electrons. The system of conducting electrons in a superconducting metal forms pairs of bound electrons ("Cooper" pairs). There is still the fundamental problem of the mechanism of superconductivity, I,e. the origin of the pairing should be explained. Indeed, pairing means that there is an attraction between the paired electrons; as a result, they can form a bound state. What is the origin of such a force? As was shown in the BCS theory, and later supported by experimental and theoretical studies of many superconducting materials, this attraction is provided by the electron-lattice interaction.

According to the quantum theory of solids, the lattice excitations in bulk metals which correspond to small ionic vibrations (a/d<<1, where a is the amplitude of vibrations and d is the lattice period), can be described as acoustic quanta (phonons) with energies $\varepsilon_{ph}^{i} = \hbar\Omega_i(\vec{q}); \vec{q} = \hbar\vec{k}$ (in the following we *set* $\hbar$ =1), with momentum $\vec{q}$ and wave number $\vec{k}$ ( k=2π/λ, where λ is the phonon's wavelength), and with i corresponding to the various phonon branches ( longitudinal, transverse, optical). For such systems the electron-lattice interaction, e.g., the energy exchange between the electrons and lattice ,can be described as radiation and adsorption of phonons and is denotedd as the electron-phonon interaction.

In the Debye model ( see, e.g., Landau and Lifshitz, 1969), all acoustic branches are described by the linear law: Ω=uq, where u is the average sound velocity. The value of the so-called Debye frequency ,which is the maximum frequency of vibrations ($\Omega_D \equiv \Omega_{max}$), is determined by the condition that the total number of vibrations $V\Omega_{max}^{3}/2\pi^{2}u^{3}$ is equal to the total number of vibrational degrees of freedom 3N (N is the number of ions). One can estimate: $\Omega_D \approx uq_{max} \approx u/d$, where d is the lattice period.

According to the BCS theory of superconductivity , pairing is provided by the electron-phonon interaction, or more specifically, by the exchange of phonons between the electrons forming the pair. This exchange means the emission of a phonon by an electron moving through the lattice and the subsequent absorption of the phonon by another electron.

In 1986, the 75th anniversary of superconductivity was marked by the discovery of a new class of superconducting materials, namely, high $T_c$ copper oxides (usually called cuprates). Bednorz and Mueller (1986) discovered that the $La_{1.85}Ba_{0.15}CuO_4$ compound became superconducting with a critical temperature $T_c \approx 30$ K , which noticeably exceeded the previous record ($T_c \approx 23.2$ K for $Nb_3Ge$). Optimization of the synthesis of the similar compound ( La-Sr-Cu-O) moved the transition temperature close to 40 K.  This achievement was quickly followed by discoveries of other high $T_c$ copper oxides (cuprates). The most studied is the $YBa_2Cu_3O_{6+x}$ (YBCO) compound with $T_c \approx 93$ K  at $x \approx 0.9$  ( Wu et al. ,1987). At  present, the highest observed value of $T_c$ is about 150 K and is for the $HgBa_2Ca_2Cu_3O_{8+x}$ compound under pressure. The discoveries of new cuprates were accompanied by intensive studies of their structure and properties (see , e.g., reviews by Kresin and Wolf, 1990; Ginsberg,

Ed.,1994). It turns out that all cuprates have a layered structure. The main structural unit that is typical for the whole family is the Cu-O plane (see Fig. 1). One should distinguish between the main layer (Cu-O plane) where pairing originates and the charge reservoir. For example, in addition to the Cu-O planes, the YBCO compound contains Cu-O chains, and the change in the oxygen content in the chain layers leads to charge transfer between these two subsystems.The charge transfer occurs through the apical oxygen ion located between the planes and chains (Fig.1).

Another important property of these novel superconductors is that they are doped materials. The doping is provided either by chemical substitution( e.g., by the La->Sr substitution in the $La_{2-x}Sr_xCuO_4$ compound; the value of $T_c \approx 40$ K corresponds to $x \approx 0.15$), or by changing the oxygen content. Doping leads to the appearance of carriers in the Cu-O planes. There are two types of carriers (see, e.g., Ashcroft and Mermin,1976). One of them (electrons) is created by the dopants which are called donors. The second type (holes; they have a positive charge) is produced by doping , which removes electrons. Some of the cuprates (e.g., Nd-Ce-Cu-O) contain electrons as the carriers. Such an important material as YBCO contains carriers that are holes. It is important that the value of $T_c$ depends strongly on the in-plane carrier concentration. The undoped parent compounds are insulators. Doping leads to conductivity and then, for larger carrier concentration, to superconductivity. There is some characteristic value of the carrier concentration, $n_m$, which corresponds to the maximum value of $T_c \equiv T_c^{max}$. The underdoped ($n<n_m$) and overdoped ($n>n_m$) regions are characterized by values of $T_c$ lower than $T_c^{max}$.

Since the discovery of high $T_c$ oxides , there has been an intensive and fruitful study of these novel materials . However, despite intensive research, the question of the mechanism for these materials is still open. There has been growing evidence, mainly from various experimental studies, that the electron-lattice  interaction is important for understanding the nature of high $T_c$ superconductivity in the cuprates. This interaction provides a direct contribution to the pairing of electrons, and also is clearly manifested in polaronic effects.  The polaronic effects appear as a result of the strong electron-lattice interaction. In this case, a moving electron polarizes the lattice, and  a shift in positions of neighboring ions forms a potential "box" for the electron. A polaron is a unit containing an electron that is moving with the lattice polarization caused by the electron itself (see,e.g., Ashcroft and Mermin,1976; Devresee, 2005,  and see Sec.IV).

This Colloquium is not a review, but rather a systematic description of our view, reflected in many publications on the subject. This Colloquium also contains an extensive list of references that are related to this subject. Our viewpoint  is that the electron-lattice interaction is an important ingredient of the current scenario and can explain superconductivity in novel systems including high  temperature superconducting cuprates.

The structure of the paper is as follows.The  general properties of superconductivity  caused by the electron- phonon interaction are discussed in Sec..II. Experimental data demonstrating the impact of this interaction are described in Sec.III. Section IV is concerned with the polaronic effect and the isotopic substitution. The phonon–plasmon mechanism for layered systems is described in Sec.V. Section VI contains a discussion of the electron-lattice interaction in the "pseudogap" state.  A critical experiment to provide more insight into the mechanism is proposed in Sec. VII. High $T_c$ superconductivity  in nanoclusters caused by the electron-vibrational interaction is discussed in Sec. VIII. Section  IX contains concluding remarks.

II.Electron-lattice interaction and the upper limit of $T_c$

The electron-lattice interaction can, in principle, lead to high values of $T_c$ (see below). Of course, this statement alone does not provide the answer to the question about the nature of the superconducting state in the cuprates, but it means that the electron-lattice interaction cannot be summarily ruled out as a potential mechanism.

One should note that immediately after the discovery of high $T_c$ oxides ,many excluded the electron-lattice interaction from the list of potential mechanisms. For the most part , this was done because of the natural temptation to introduce something new and exciting into the field as opposed to relying on the important principle of Occam's razor  "*pluralitas non est ponenda sine neccesitate*" ( "one should not increase, beyond what is necessary, the number of entities required to explain anything"). An additional key factor was the conviction that despite the electron-phonon interaction being  successful as an explanation for superconductivity in conventional materials, this mechanism is not sufficient to explain the observed high values of $T_c$. We address this important second aspect.

The Bardeen-Cooper-Schrieffer theory (Bardeen et al. ,1957) was developed in the weak coupling approximation ($\lambda \ll 1$, where $\lambda$ is the electron-phonon coupling constant at T=0; its value reflects the strength of the electron-lattice interaction). Electron-phonon coupling leads to attraction between electrons in a superconductor. More specifically, an electron polarizes the lattice, that is,it induces ionic motion which affects another electron. As mentioned, In the quantum picture the process can be visualized as an exchange of phonons; such an exchange leads to an attraction between electrons [ a detailed description of this interaction has been given , e.g., by Ashcroft and Mermin (1976) and Kresin and Wolf(1990)]. In superconductors, this attraction overcomes Coulomb electron repulsion.

The expression for the critical temperature derived in the BCS theory has the form

$$k_B T_c \approx \tilde{\Omega} e^{-1/(\lambda - \mu^*)} \qquad (2.1)$$

where $\tilde{\Omega}$ is the characteristic phonon frequency, $\tilde{\Omega} \approx \Omega_D$, $\Omega_D$ is the Debye frequency, and $\mu^* = V_c \left[ 1 + V_c \ln(\varepsilon_0 / \tilde{\Omega}) \right]^{-1}$ describes the Coulomb repulsion; $\varepsilon_0 \approx E_F$, where $E_F$ is the Fermi energy; usually $\mu^* \approx 0.1$ .

As mentioned, Eq. (2.1) is valid in a weak coupling approximation. Since $\lambda << 1$ (e.g., $0.5 \gtrsim \lambda$), one could easily come to the conclusion, which follows from Eq. (2.1), that $T_c$ should be at least of an order of magnitude below the Debye temperature (the Debye temperature $\theta_D$ is determined by the relation: $k_B \theta_D = \hbar \Omega_D$; in the following discussion , we have set $k_B = \hbar = 1$ so that energy E, frequency $\Omega$, and temperature T all have the same units).

In many superconductors, the condition $\lambda \ll 1$ is not satisfied and $\lambda \gtrsim 1$. For example, in lead, $\lambda=1.4$; in mercury, $\lambda=1.6$; and in the alloy $Pb_{0.65}Bi_{0.35}$ ,the coupling constant has the value $\lambda \approx 2.1$ (see, e.g., Allen and Dynes (1975); Wolf (1985)). To understand the consequences of these high values of $\lambda$ , it is necessary to go beyond the limit of weak coupling. This more universal approach was developed shortly after the creation of the BCS theory (Eliashberg, 1961, 1963) and allows us to analyze the properties of superconductors with strong electron-phonon coupling.

Strong coupling theory is a generalization of the theory of normal metals (Migdal ,1960). It is also based on the method developed by Gor'kov (1958) which was initially applied for the weak coupling case (see ,e.g., Abrikosov et al.,1963). A detailed description of the fundamentals of superconductivity with strong coupling can be found in a number of reviews and monographs (see, e.g., Scalapino ,1969; Grimvall ,1981; Kresin et al.,1993) and is based on the Green's function method of the many-body theory.

We introduce here the main quantities that enter the theory. The phonon spectrum contains a continuous distribution of phonon frequencies and it is described by the phonon density of states $F(\Omega)$, where $\Omega$ is the phonon frequency. An important material dependent parameter is $\alpha^2(\Omega)F(\Omega)$ where $\alpha^2(\Omega)$ is a measure of the phonon frequency dependent electron-phonon interaction. The electron-phonon coupling constant $\lambda$, which determines the value of $T_c$ [(see Eq.(2.1) and Eqs.(2.8)-(2.12)] can be written as

$$\lambda = 2\int \alpha^2(\Omega)F(\Omega)\Omega^{-1}d\Omega \qquad (2.2)$$

One can introduce the characteristic phonon frequency $\tilde{\Omega}$, which is defined as an average over $\alpha^2(\Omega)F(\Omega)$,

$$\tilde{\Omega} = <\Omega^2>^{1/2}. \qquad (2.3)$$

The average is determined by $\langle f(\Omega)\rangle = \dfrac{2}{\lambda}\int d\Omega f(\Omega)\alpha^2(\Omega)F(\Omega)\Omega^{-1}$, so that

$<\Omega^2> = \dfrac{2}{\lambda}\int \Omega\alpha^2(\Omega)F(\Omega)d\Omega$; the coupling constant is defined by Eq.(2.2).

The main quantity of interest is the pairing order parameter $\Delta(\omega)$. The pairing energy gap can be determined as the root of the equation $\omega=\Delta(i\omega)$.

The equation for the pairing order parameter $\Delta(\omega)$ has the form (at T=0K):

$$\Delta(\omega) = \left[Z(\omega)\right]^{-1}\int\limits_{0}^{\omega_o} d\omega' P(\omega')\left[K_+(\omega,\omega') - \mu*\right] \qquad (2.4)$$

where

$$[1 - Z(\omega)]\omega = \int_0^\infty d\omega' N(\omega') K_-(\omega, \omega')$$

$$P(\omega) = \mathrm{Re}\{\Delta(\omega)[\omega^2 - \Delta^2(\omega)]^{-1/2}\}$$

$$N(\omega) = \mathrm{Re}\{|\omega|[\omega^2 - \Delta^2(\omega)]^{-1/2}\}$$

$$and$$

$$K_\pm(\omega, \omega') = \int d\Omega \alpha^2(\Omega) F(\Omega) \left( \frac{1}{\omega' + \omega + \Omega + i\delta} \pm \frac{1}{\omega' - \omega + \Omega - i\delta} \right).$$

Here $\Omega$ is the phonon frequency and $Z$ is the so-called renormalization function describing the "dressing" of electrons moving through the lattice. Equations (2.1) and (2.2) also contain the Coulomb pseudopotential $\mu^*$. The important aspect of pairing is the logarithmic weakening of the Coulomb repulsion (see Bogoliubov et al. ,1959; Khalatnikov and Abrikosov,1959; Morel and Anderson,1962) , which is related to the difference in the energy scales of the attractive and repulsive effects [see discussion following Eq.(2.1)]. The attraction is important in an energy interval $\tilde{\Omega}$, whereas the repulsion is characterized by the energy scale $\varepsilon_0 \sim E_F$, where $E_F$ is the Fermi energy. In usual metals $E_F \approx 10$ eV, the characteristic phonon frequency $\tilde{\Omega} \approx$ 20-50 meV, so that $E_F >> \tilde{\Omega}$. As a result, the Coulomb pseudopotential $\mu^* = V_c [1 + V_c \ln(\varepsilon_0 / \tilde{\Omega})]^{-1}$ contains a large logarithmic factor that reduces the contribution of the Coulomb repulsion. For the cuprates, the electronic energy scale $\varepsilon_0 \sim 1$ eV, and although it is smaller than the corresponding energy scale in conventional superconductors it is still much larger than the scale of the lattice energy. For simplicity, we omit $\mu^*$ below in some equations.

At finite temperature it is convenient to use the thermodynamic Green's function formalism (see, e.g., Abrikosov et al.,1963). Then the major equation can be written in the form

$$\Delta(\omega_n)Z = T\sum_{\omega_{n'}}\int d\Omega\,\Omega^{-1}\alpha^2(\Omega)F(\Omega)D(\omega_n - \omega_{n'};\Omega)F^+(\omega_{n'})$$

(2.5)

*where*

$$D = \Omega^2\left[\left(\omega_n - \omega_{n'}\right)^2 + \Omega^2\right]^{-1}$$

is the so-called phonon Green's function. Eq.(2.5) can be approximated to a high degree of accuracy using Eq. (2.2),

$$\Delta(\omega_n)Z = \lambda T\sum_{\omega_{n'}}D(\omega_n - \omega_{n'};\tilde{\Omega})F^+(\omega_{n'})$$
,

(2.6)

where $\tilde{\Omega}$ is the characteristic phonon frequency [ see Eq.(2.3)], the coupling constant is defined by Eq. (2.2), and

$$F^+ = \Delta(\omega_n)/\left(\omega_n^2 + \xi^2 + \Delta^2(\omega_n)\right)$$

is the pairing Green's function, introduced by Gor'kov (1958); $\xi$ is the electron energy referred to the chemical potential. One can also write out the equation for the renormalization function Z.

McMillan (1968) introduced a convenient expression for the coupling constant $\lambda$,

$$\lambda = \nu \left\langle I^2 \right\rangle / M \tilde{\Omega}^2 \ . \tag{2.7}$$

In Eq. (2.7) $\nu = m^* p_F / 2\pi^2$ is the bulk density of states, $\left\langle I^2 \right\rangle$ contains the average value of the electron-phonon matrix element I (see, e.g., Grimvall,1981), and $\tilde{\Omega}$ is defined by Eq. (2.3). One can see from Eq. (2.7) that $\lambda$ is not a universal constant, but a material-dependent parameter.

Equation (2.6) is especially convenient for evaluating $T_c$ and for analyzing the thermodynamic properties. It is important that the strong coupling superconductivity theory is valid if $\tilde{\Omega} << E_F$. This is the **only** condition for its applicability. It is important to stress also that there is no limit on the value of $T_c$, and the theory even allows $T_c$ to exceed the Debye temperature.

As noted above, the derivation of Eqs. (2.2) and (2.5) is based on a special method (Green's function formalism), and its description is beyond the scope of this paper. It is worth noting that these equations are a generalization of the BCS theory. Indeed, if we assume that the electron-phonon coupling is weak (then $T_c << \bar{\Omega}$), one can neglect the dependence of the D function on $\omega_n = (2n+1)\pi T_c$, and then $\Delta(\omega_n)$=const. At $T=T_c$ one should put $\Delta$=0 in the denominator of Eq. (2.6). Then one can calculate $T_c$. Performing a summation, we arrive at the usual BCS expression (2.1).

However, strong coupling theory is based on one very important assumption. Namely, it assumes that the phonon spectrum is fixed, and this implies that the lattice is not affected by the pairing. Strictly speaking, this is not the case, and if the value of the coupling constant exceeds some value $\lambda_{max}$, then the lattice could become unstable. This problem was studied by Browman and Kagan (1967), and by Geilikman (1971, 1975) . Based on rigorous adiabatic theory, one can prove that the change in phonon characteristic frequency caused by the electron-lattice interaction is small, and the lattice becomes unstable (that is, the characteristic frequency becomes imaginary) only at very large values of $\lambda$ ($\lambda >> 1$) . Therefore, high values of $T_c$ are theoretically possible within this framework.

It is interesting that an explicit expression for $T_c$ depends on the strength of the coupling. As noted above, the BCS expression (2.1) is valid for weak coupling superconductors only. For larger values of $\lambda$ ($1.5 \gtrsim \lambda \gtrsim 1$) one can use the expression obtained by McMillan (1968) and then modified by Dynes (1972). This expression has the form

$$T_c = \frac{\tilde{\Omega}}{1.2} \exp\left[-\frac{1.04(1+\lambda)}{\lambda - \mu^*(1+0.62\lambda)}\right] \qquad (2.8)$$

If the coupling constant $\lambda$ is large ($\lambda > 1.5$), one should use a different expression for the critical temperature. We initially discuss the case of very strong coupling ($\lambda \gtrsim 5$; then $\pi T_c \gtrsim \tilde{\Omega}$). In this case, the dependence of $T_c$ on $\lambda$ differs drastically from the dependences given by Eqs. (2.1) and (2.8). As shown initially by Allen and Dynes (1975) using numerical calculations, and later analytically by Kresin et al. (1984), this dependence has the form (here we assume $\mu^*=0$)

$$T_c = 0.18\lambda^{1/2}\hat{\Omega} \qquad (2.9)$$

Kresin et al. (1984) also obtained the expression for $T_c$, when $\mu^* \neq 0$, that is,

$$T_c = 0.18\lambda_{eff}^{1/2}\tilde{\Omega}$$
$$\lambda_{eff} = \lambda(1+2.6\mu^*)^{-1} \qquad (2.10)$$

This analytical expression was obtained using the matrix method (Owen and Scalapino, 1971). The scaling behavior for $T_c$ can be seen directly from Eq. (2.6). Indeed, if $\pi T_c \gg \tilde{\Omega}$, then one can neglect $\tilde{\Omega}^2$ in the denominator of the phonon Green's function, and then one can then directly see the scaling behavior $T_c \propto \lambda^{1/2}\tilde{\Omega}$.

One can see from Eqs. (2.9) and (2.10) that for large $\lambda$ the expression for $T_c$ is very different from Eqs. (2.1) and (2.8). As mentioned earlier, Eq.

(2.10) is valid for $\lambda \gtrsim 5$. For the intermediate case, one can use the general equation (Kresin, 1987a) that was obtained by solving Eq. (2.6) and is valid for any value of the coupling constant,

$$T_c = \frac{0.25\tilde{\Omega}}{(e^{2/\lambda_{eff}} - 1)^{1/2}}$$
$$\lambda_{eff} = (\lambda - \mu^*)\left[1 + 2\mu^* + \lambda\mu^* t(\lambda)\right]^{-1} \tag{2.11}$$

The universal function $t(\lambda)$ decreases exponentially with increasing $\lambda$; $t(\lambda)$ can be approximated quite accurately by $t(\lambda) = 1.5\exp(-0.28\lambda)$; such an approximation was proposed by Tewari and Gumber (1990), see also Kresin and Wolf (1990). If we neglect $\mu^*$, we obtain

$$T_c = \frac{0.25\tilde{\Omega}}{\left(e^{2/\lambda} - 1\right)^{1/2}} \tag{2.12}$$

As mentioned above, Eqs. (2.11) and (2.12) are valid for any strength of the coupling. One can easily see that for the weak coupling case Eq. (2.11) reduces to Eq. (2.1), whereas for $\lambda \gg 1$ we obtain the dependence of Eq. (2.10).

Equation (2.12) was obtained by Kresin (1987a) as a solution of Eq. (2.6). The dependence of Eq. (2.12) was used as a trial function, and then it was demonstrated that it satisfied Eq.(2.6) with a high degree of precision. Later the same expression was obtained analytically by Bourne et al. (1987)

for the model case: $\alpha^2(\Omega)F(\Omega)$=const for $0<\Omega<\Omega_{max}$ and $\alpha^2(\Omega)F(\Omega)=0$ for $\Omega>\Omega_{max}$.

One can see directly from Eqs. (2.9) - (2.11) that there is a large range of values for the coupling constant where the lattice is still stable, and the value of $T_c$ is high. In principle, $T_c$ can reach room temperature (e.g., for $\lambda_{eff}\approx5$; $\tilde{\Omega}\approx60$ meV). The values of $T_c$ observed in the cuprates are even more realistic (e.g., for $\lambda_{eff}\approx3$-3.5, $\tilde{\Omega}\approx25$ meV)

The question about an upper limit of $T_c$ for the phonon mechanism has some interesting history. Based on the so-called Froelich Hamiltonian ,which is the sum of the electronic term, the phonon term with experimentally measured phonon frequency, and the electron-phonon interaction, one can obtain

$$\Omega = \Omega_0(1 - 2\lambda)^{1/2} \qquad\qquad (2.13)$$

(Migdal, 1960). Based on this expression, one can conclude that the value of the coupling constant $\lambda$ cannot exceed $\lambda_{max}$=0.5, and this implies that the value $T_c \lesssim 0.1\,\tilde{\Omega}$ $(\tilde{\Omega}\approx\Omega_D)$ is the upper limit of $T_c$. Indeed, such a point of view was almost generally accepted after the appearance of the BCS theory. However, it soon became clear that something is wrong with this criterion, since there were many superconductors discovered with $\lambda>0.5$ (e.g., Sn, Pb,Hg). The problem was clarified later by Browman and Kagan, (1967), and by Geilikman (1971, 1975). As mentioned, the total Hamiltonian that leads to Eq. (2.13) contains terms describing free electronic and phonon fields and their interaction; the phonon term contains an experimentally observed phonon spectrum, including an acoustic branch. However, one can demonstrate that the formation of an acoustic dispersion law is also provided by the electron-ion interaction. In other words, in this model we are double counting. This means that the analysis of the electron-phonon interaction has to be carried out with considerable care. This has been done, based directly on the adiabatic approximation by Geilikman (1971, 1975) and Browman and Kagan (1967), see also, the review: Kresin et al. (1993). The theory starts from the initial picture of electrons and ions, and the formation of the phonon branch and the residual electron-phonon interaction has been obtained by rigorous and self-consistent analysis. The conclusion is that the electron-phonon interaction does **not** lead to the dependence of Eq. (2.13).

Note that this conclusion does not mean the absence of lattice instabilities. In fact, the electron-phonon interaction can lead to various instabilities, especially for systems containing low-dimensional units. But this fact does not support the conclusion about the existence of an upper value of $\lambda$ and, correspondingly, an upper limit for $T_c$. It is likely that such a limit exists for very large $\lambda$ ($\lambda \gtrsim 10$), but this is still an open question.

Another faulty restriction on $T_c$ was later proposed and was based on the McMillan equation (2.8). Indeed, this equation taken at face value leads to an upper limit of $T_c$. If one neglects $\mu^*$ for simplicity, one can easily find that the maximum value of $T_c$ corresponds to $\lambda=2$; then $T_c^{\max} \approx \tilde{\Omega}/6$. This conclusion, however, assumes that Eq. (2.8) is valid for $\lambda > 1.5$. But the McMillan equation is valid only for $\lambda \lesssim 1.5$. Therefore, the value $\lambda=2$ is outside of the range of its applicability.

The treatment based on Eqs. (2.10) and (2.11) leads to a very different conclusion, namely to the absence of the upper limit for $T_c$. As noted before, experimentally large values of $\lambda$ have been determined for several superconductors. For example, $\lambda \approx 2.1$ for $Pb_{0.65}Bi$, $\lambda \approx 2.6$ for Am-$Pb_{0.45}Bi_{0.55}$ (Allen and Dynes,1975: see also the review: Wolf,1985) . Also, if the material is characterized by relatively large values of **both** $\lambda$ and $\tilde{\Omega}$, it might have a very high value of $T_c$.

As stressed above, this conclusion by itself does not mean that high temperature superconductivity in the cuprates is provided by the electron-phonon interaction. This can be determined only by special and detailed experimental study (e.g., by tunneling spectroscopy), but such a mechanism cannot be ruled out on any theoretical grounds.

## III. **Experimental data and analysis**

Only some selected experimental techniques can provide information about the pairing mechanism. Indeed, many experimental studies are not sensitive to the pairing interaction. For example, thermodynamic and electromagnetic properties contain the energy gap $\varepsilon_0$ as a parameter and, since the energy gap is directly proportional to the critical temperature (according to the BCS theory $\varepsilon_0 = 1.76 k_B T_c$), $T_c$ becomes the key parameter of the theory. As a result, all such properties are parametrized by the critical temperature, and they are not sensitive to the nature of the pairing interaction. Only selected methods are sensitive to the nature of the interaction which provides the observed values of $T_c$.

According to the BCS theory, the pairing is provided by the electron-phonon interaction, that is, by phonon exchange between the paired electrons. However, after Little's paper (1964) it become clear that the pairing can be caused by other excitations as well. Among these excitations are electronic ones. This electronic mechanism can be important if the material contains two groups of electrons. Excitations within one of these groups serve as "agents" giving rise to pairing in the other group. Another electronic mechanism represents exchange through coupling to plasmons which are electronic collective excitations (see Sec.V). Pairing can be provided also by exchange of magnetic excitations (magnons).

In principle, these and other mechanisms can provide pairing in novel materials. In addition, the superconducting state can be caused by the contributions of different excitations. Based on special experiments, one should be able to determine the key factors responsible for pairing in these novel materials. Below we discuss some of these techniques and their relevance to the problem of determining the mechanism of pairing in the cuprates.

A.   Tunneling spectroscopy

1. McMillan-Rowell method

It is only because of the special tunneling method developed by McMillan and Rowell (1965, 1969) that we have rigorous evidence that the phonon mechanism, that is, the mechanism based on the electron-phonon coupling is the dominant one for conventional superconductors.

Superconducting tunneling spectroscopy was developed by McMillan and Rowell and described in their review (McMillan and Rowell,1969), see also the review by Wolf (1985). Here are some key elements of this approach.

Tunneling spectroscopy (see,e.g., Burstein and Lindqvist, 1969) is based on observation of the tunneling of electrons through a typically very thin ($\approx$10 A) insulating barrier separating the superconductor that is being studied and some other metallic layer. One then measures the tunneling current as a function of the applied voltage and analyzes the result according to the McMillan-Rowell procedure (McMillan and Rowell, 1969). There are several experimental techniques that have been used to generate tunneling spectra. The most widely used method requires the deposition of the superconducting electrode, the formation of a barrier, by either oxidation of the superconductor or depositing an insulating layer , and then final deposition of another metallic or superconducting electrode on top of the insulator.

The important quantities that need to be measured are the direct current-voltage characteristic I-V, the derivative of the I-V, dI/dV as a function of the voltage , and the second derivative $d^2I/dV^2$ also as a function of the voltage. These data must be taken with the sample in both the normal and the superconducting state. If the counter electrode is a normal metal, then measurements of the normalized conductance of the junction $\sigma = (dj/dV)_S/(dj/dV)_n$ allow us to determine the key quantity, the tunneling density of states $N_T(\omega)$,

$$N_T(\omega) = \text{Re}[|\omega|/(\omega^2 - \Delta^2)^{1/2}], \qquad (3.1)$$

where $N_T(\omega)$ is normalized by the density of states in the normal state. Note that the tunneling density of states contains the order parameter $\Delta \equiv \Delta(\omega)$. This analysis is based on Eq. (2.4), i.e., it is assumed that the superconducting state is provided by the electron-phonon interaction. It is important to note also that the peak position in the function $\alpha^2(\Omega)F(\Omega)$ corresponds to the point of the most negative slope in the tunneling conductance. Therefore, the second derivative $d^2I/dV^2$ allows one to locate the peaks in the phonon spectrum.

The key part of the method is the inversion procedure. The function $\Delta(\omega)$ determined from the tunneling conductance measurements, can be used to evaluate the function $\alpha^2(\Omega)F(\Omega)$ and the Coulomb pseudopotential $\mu^*$ from Eqs. (3.1) and (2.4). Inverting eq. (2.4) allows one to determine the function $\alpha^2(\Omega)F(\Omega)$ introduced in Sec. II, and $\mu^*$. The McMillan-Rowell method involves numerically solving the integral equations (2.4) for a given set of parameters, calculating $N_T(\omega)$ [ (Eq.3.1)], comparing the calculated values to the measured values, adjusting the input parameters, and iterating the procedure until the calculated tunneling density of states matches the measured one. As mentioned above, a detailed description of the procedure and the application to Pb can be found in McMillan and Rowell (1969); see also Wolf (1985).

The function $\alpha^2(\Omega)F(\Omega)$ contains two factors. One of them $[\alpha^2(\Omega)]$, depends weakly on frequency, whereas the phonon density of states $F(\Omega)$ usually contains two peaks, corresponding to transverse and longitudinal phonons. The peaks occur in the short -wavelength region, which makes a major contribution to pairing; the dispersion law in this region deviates from the usual acoustic law dependence and is close to being rather flat; this leads directly to a peaked structure of $F(\Omega) \propto dq/d\Omega$ ). A further important check on this procedure can be provided using inelastic neutron scattering measurements to determine the phonon density of states $F(\Omega)$. These measurements are not related to superconductivity. Comparison of the tunneling and neutron scattering measurements can provide important information. In fact, one can compare the position of the peaks determined by these two different methods. The coincidence of these positions verifies the initial assumption that superconductivity in the material of interest is caused by the electron-phonon coupling, that is, pairing occurs by exchange of phonons. The dependence $\alpha^2(\Omega)F(\Omega)$ and the value of $\mu^*$ obtained by inverting the tunneling spectrum can be used to calculate $T_c$ directly from Eqs. (2.4) and (2.6), or with the use of Eq. (2.3) , and then Eqs. (2.8) and (2.11) ,which can then be compared with the experimental value.

This method was applied to many conventional superconducting elements (see Fig. 2) and compounds, and by virtue of the remarkable agreement between theory and experiment, the mechanism in most conventional superconductors has been proven to be the electron-phonon interaction, or as is usually stated the phonon mechanism..

## 2.Tunneling studies of the cuprates

It is very temping to use tunneling spectroscopy to study the nature of high $T_c$ superconductivity in the cuprates. However, there is a serious challenge. As we know the coherence length , which is defined as $\xi = \hbar v_F / 2\pi T_c$ (where $v_F$ is the Fermi velocity),  is an  important parameter; its value characterizes the scale of pairing and can be visualized as the size of the pair. For usual superconductors,  the value of  $\xi$  is rather large  ($\sim 10^3$-$10^4$ A), whereas for the cuprates it is quite small: $\xi \approx$ 15-20 A. The length scale for providing the tunneling current at the interface between the superconductor and the insulator, that is, the depth over which the tunneling current originates is the pairing coherence length, and as noted above for conventional superconductors this is a large quantity that greatly exceeds the thickness of the surface layer. In the cuprates, the coherence length is very short, and this makes the measurements difficult. Nevertheless, such experiments were performed.

One of the first tunneling experiments (Dynes et al.,1992) was carried out to study the yttrium- barium- copper oxide (YBCO) compound. Unfortunately, this paper has stayed mainly unnoticed by the high $T_c$ community. The inversion procedure carried out in this paper resulted in the dependence $\alpha^2(\Omega)F(\Omega)$ , shown in Fig.3 . The calculated value of the critical temperature was $T_c \approx 60$ K. This value lies below the experimental one, but is still quite high. In addition, the experimentally measured (via neutron scattering) peak in the phonon density of states is somewhat below the peak position for the function $\alpha^2(\Omega)F(\Omega)$ obtained from the inversion procedure. Such a difference might reflect the presence of some additional mechanism, or perhaps is caused by a pair-breaking effect (Abrikosov and Gor'kov,1961; Kresin and Wolf, 1995) that has not been considered in the analysis. We believe that this effect is caused by magnetic impurities. As we know, the pair is formed by two electrons with opposite momenta and opposite spin. Each localized magnetic moment (magnetic impurity) is trying to align the spins in the same direction, and this destroys the pairing. The presence of such broken pairs leads to the appearance of a gapless spectrum. Indeed, YBCO, contrary to conventional materials, does not display a sharp gap structure; its spectrum is rather gapless. In connection with this, it is interesting to note that Dynes et al. (1992), by applying an external magnetic field induced gaplessness in Pb. They carried out an analysis using the inversion procedure and observed a result (with proper scaling) very similar to that observed for YBCO.

Break junction tunneling spectroscopy ,which provides a high quality contact, was employed by Aminov et al. (1994) and Ponomarev et al. (1999). They demonstrated (Fig. 4 ) that the current-voltage characteristic for the $Bi_2$ $Sr_2Ca$ $Ca_2$ $O_8$ compound contains an additional substructure that strongly correlates with the phonon density of states; phonon density of states was obtained by Renker et al. (1987;1989) using inelastic neutron scattering . Such a correlation is a strong indication of the importance of the electron-phonon interaction.

The tunneling conductance of $Bi_2$ $Sr_2Ca$ $Ca_2$ $O_8$ was measured by Shiina et al.(1995), Shimada et al. (1998), and Tsuda et al.(2007). They also observed a correspondence of the peaks in $d^2I/dV^2$ and the phonon density of states. Moreover, the McMillan-Rowell inversion was performed, and the result was supportive of the electron-phonon scenario. The spectral function $\alpha^2(\Omega)F(\Omega)$ contains two groups of peaks at $\Omega \approx 15\text{-}20$ and $\approx 30\text{-}40$ meV. The positions of the peaks corresponds with a high degree of accuracy to the structure of the phonon density of states. The coupling constant $\lambda$ appears to be equal to about 3.5; such strong coupling is sufficient [see Eq. (2.11)] to provide the observed value of $T_c$ ($T_c \approx 90$ K).

Large values of $2\Delta/T_c \gtrsim 10$ ,which greatly exceed the conventional values, observed for the underdoped sample (Miyakawa et al.,(2002) can be explained by the fact that the energy gap persists above $T_c$ up to $T_c^*$ ("pseudogap' region) and is in agreement with recent data (Gomes et al.,2007). This effect is caused by an intrinsic inhomogeneity of the sample (see Sec. VI and the review by Kresin et al., 2006).

Another tunneling technique that appears to be a powerful tool in many studies is scanning tunneling microscopy (STM). This method is widely used in order to obtain information about the local structure of the order parameter, its inhomogeneity, etc. This type of tunneling (STM) can also be used to perform a study that can probe the mechanism of high $T_c$. For example, Lee et al. (2006) carried out an STM analysis of the $Bi_2Sr_2CaCu_2O_{8+\delta}$ compound. As a part of the study, measurements of the tunneling current and its second derivative $d^2I/dV^2$ were performed. The locations of the peaks in the second derivative coincide with the position of specific phonon modes. This is a strong indication of the importance of electron-phonon coupling. Of course, a complete analysis requires the inversion procedure ,which so far has not been carried out. Lee et al. stated that they are planning to perform this procedure ; perhaps, soon it will be done. Recently a similar correlation between the tunneling and Raman data for La-Sr-Cu-O was observed by Shim et al. (2008).

A detailed STM study of the three-layer $Bi_2Sr_2CaCu_2O_{10+\delta}$ compound was performed recently by Levy de Castro et al. (2008). They concluded that it is necessary to take into account the band structure of the material ,and especially the presence of the van Hove singularity, that is, the cusp in the electronic density of states that often appears in compounds with lower- dimensional substructural units (e.g., planes and/or chains). The interaction with some collective mode is also an essential factor in the analysis. These factors allow us to describe the observed features in the conductivity such as the dip asymmetry as well as the observed dip-hump structure (Renner and Fischer, 1995). Phonons could provide such a collective mode, but some magnetic excitations could do the same. Additional measurements can determine the exact nature of the mode.

As a whole, tunneling spectroscopy continues to be a powerful and promising tool.

B.Infrared spectroscopy.

A new method based on precise infrared measurements can be used to reconstruct the function $\alpha^2(\Omega)F(\Omega)$. This method was proposed by Little and collaborators [ for a description see Little et al, 1999 )] and is based on the so-called thermal-difference-reflectance spectroscopy . This method was demonstrated by Holcomb et al. (1993,1994,1996) and allows one to determine the function $\alpha^2(\Omega)F(\Omega)$ for an energy interval that is larger than that in the tunneling method. The reflectivity of the sample was measured with a high degree of precision at different temperatures, and the ratio of the difference relative to their sum was determined. The theoretical method developed by Shaw and Swihart (1968) was used in order to perform the inversion for thallium cuprate (2212) samples. The larger extent of the accessible energy range Little et al.,(1999) to take into consideration the electronic modes whose energy lies noticeably higher than typical phonon energies.

As we know, Little (1964) introduced the electronic mechanism of superconductivity in his pioneering work ;see also the review by Gutfreund and Little (1979). Many interesting and novel aspects of various electronic mechanisms were also described by Ginzburg (1965), see also Ginzburg et al. (1982) and  Geilikman (1965, 1973). In these papers ,pairing is provided not by phonons but by electronic excitations, e.g., by excitons [ while a usual electronic excitation corresponds to an appearance of an electron at $E>E_F$ and a hole at $E<E_F$,  exciton can be viewed as a bound electron-hole state; see, e.g., Yu and Cardona (1999). It is important to note that the superconducting state can benefit from the large energy scale characteristic of the electronic mechanism.

According to Little et al. (2007), the superconducting state in the cuprates is caused by both phonon and electronic contributions , and each of them is of key importance. The phonon contribution is characterized by an intermediate coupling constant ($\lambda_{phon} \approx 0.9$). In addition, there are two electronic peaks at higher energies with the strengths: $\lambda_{1.2\ eV} \approx 0.1$ and $\lambda_{1.7\ eV} \approx 0.3$. This combination can provide the observed high values of $T_c$. The excitonic–like excitations, namely the d-d transitions of the Cu ions. are the electronic excitations of interest. Resonant Inelastic x-ray emission spectroscopy was employed to confirm the presence of such excitations. For our purpose it is important to note that although the electronic mechanism in this scenario is playing an important role, the contribution of the electron-phonon interaction is essential to obtain the high value of $T_c$ observed.

 C.Photoemission and  ultrafast electron spectroscopy.

After the discovery of the high $T_c$ cuprates,  the photoemission technique was developed as a powerful tool  used  to obtain information about the energy spectrum and electronic structure of these novel materials. Photoemission experiments indicating the presence of substantial electron-phonon coupling were published by Lanzara et al. (2001). They studied different families of hole-doped cuprates, Bi2212, LSCO, and  Pb-doped Bi2212, and they investigated the electronic quasiparticle dispersion relations. A kink in the dispersion around 50-80 meV was observed. This energy scale corresponds to the energy scale of some high energy phonons; it is much higher than the energy scale for the pairing gap. Such a kink can not be explained by the presence of a magnetic mode, because such a mode does not exist in LSCO , while the kink structure was also observed in this cuprate.

The structure observed by photoemission is consistent with the data on the phonon spectrum obtained by neutron spectroscopy. These measurements were also used in order to obtain a crude estimate of the electron-phonon coupling, since the quasiparticle velocity in the low temperature region is renormalized by the electron phonon coupling constant $\lambda$: $v = v_b(1+\lambda)^{-1}$, where $v_b$ is the bare (unrenormalized) velocity, which corresponds to the high temperature region. This estimate indicates substantial electron-phonon coupling.

A different type of spectroscopy, so-called ultrafast electron crystallography was employed by Gedik et al. (2007). The $La_2CuO_{4+\delta}$ compound was used and doping by photoexcitation was performed. It is interesting to note that the number of photon induced carriers per copper site was close to the density of chemically doped carriers in the superconducting compound. The study of time-resolved relaxation dynamics demonstrated the presence of transitions to transient states which are characterized by structural changes (noticeable expansion of the c-axis). Such a large effect on the lattice caused by electronic excitations is a strong signature of the electron-lattice interaction.

D.Isotope effect.

The isotope effect played an important role in understanding superconductivity. This effect manifests itself in the dependence of $T_c$ on the ionic mass. This dependence has the form:

$$T_c \propto M^{-\alpha} \quad , \tag{3.2}$$

where M is the ionic mass and $\alpha$ is the so-called isotope coefficient. If we neglect $\mu^*$ and consider the simplest case of a monatomic lattice, then according to Eq. (2.1) , $\alpha$=0.5, since $T_c \propto \bar{\Omega} \propto M^{-1/2}$. Note that the pure electronic or magnetic mechanisms of pairing do not involve participation of the lattice,and therefore do not contribute to the isotope effect.

The isotope effect (Maxwell, 1950; Reynolds et al., 1950) provided strong evidence that the electron-lattice interaction is involved in the formation of Cooper pairs. However, the isotope effect is a very complex phenomenon, and it is difficult to carry out a quantitative analysis that determines the degree of involvement of the lattice in the formation of the superconducting state and/or its contribution relative to other possible mechanisms. Indeed, there are many other factors that can affect the value of the isotope coefficient $\alpha$. Among them are the Coulomb pseudopotential $\mu^*$ which depends explicitly on phonon frequency. Anharmonicity of the lattice is an another factor that can lead even to negative values of $\alpha$. This situation becomes even more complicated if the material contains several varieties of ions, and this is exactly the situation for compounds and alloys (Geilikman, 1976).  Inhomogeneity of the sample, e.g., the coexistence of normal metal and superconducting regions (proximity effect), also strongly affects the isotopic dependence (Kresin et al., 1997). The presence of pair-breakers, e.g., magnetic impurities (for the D-wave case even nonmagnetic impurities act as pairbreakers) is another factor (Carbotte et al.,1991; Kresin et al., 1997). A peculiar polaronic effect can also manifest itself in an isotopic dependence; this effect will be discussed in Sec. IV.

It is interesting to note that the isotope effect has been observed in the cuprates and its temperature dependence is a peculiar one [see, e.g.,Franck et al., 1991 and the reviews by Franck ,1994, and by Keller,2005)]. More specifically, the value of $\alpha$ is relatively small at optimum doping, but increases with decreasing doping up to values that are even larger than that in the BCS theory. We discuss this feature in Sec. IV. But as described above, it is hard to draw any quantitative conclusion based solely on the value of $\alpha$.

It is interesting to note that not only $T_c$, but other quantities can also display an isotopic dependence. Among them is the penetration depth (see Sec.IV). We mentioned above (Sec. III.C) that the electron-lattice interaction manifests itself in a peculiar behavior of the phonon dispersion curve. This was detected using the photoemission technique. According to Gweon et al. (2004), the isotope substitution $O^{16}$ --> $O^{18}$ strongly affects this dispersion curve. However, the latest study by Douglass et al. ( 2007) showed a much smaller impact of the isotopic substitution , which is more consistent with the STM data by Lee et al. (2006) , see, Sec. III.A.2.

For our purposes, it is important to realize that the isotope effect strongly indicates that the ionic system and the electron-lattice interaction are involved in the formation of the superconducting state in the cuprates. As for a quantitative analysis, this should be carried out with considerable care, because there are many factors affecting the isotopic dependence. As a result, other techniques such as tunneling spectroscopy can provide more substantial information about the nature of the pairing and interplay of various contributions.

E. Heat capacity

A study of thermodynamics properties can also provide information about the pairing mechanism. This is due to the fact that the effective mass and the electronic heat capacity are renormalized by the electron-phonon interaction ( see, e.g, Grimvall,1981; Kresin and Zaitsev ,1978). Namely, $m^* = m_b[1+\lambda(T)]$, here $m^*$ and $m_b$ are the values of the effective mass and band mass, respectively. Also, the Sommerfeld constant $\gamma$ is given by

$\gamma \equiv \gamma(T) = \gamma_0[1+ 2\int d\Omega \alpha^2(\Omega)F(\Omega)\Omega^{-1}g(T/\Omega)]$, where $g(x)$ is the universal function; $g \rightarrow 0$ if $T >> \tilde{\Omega}$, and $9=1$ at T=0 K, so that $\gamma(0)= \gamma_0 (1+\lambda)$ ; see Eq. (2.2) . The presence of the second term in the expression for $\gamma(T)$ reflects the fact that moving electrons become "dressed" by the phonon cloud. As the temperature increases, the "cloud" becomes weaker, so that $\gamma(T)$ decreases. As a result, the measurements of electronic heat capacity at high temperatures and in the low temperature region can be used to evaluate the value of the electron-phonon coupling constant which determines $T_c$. Such measurements were performed by Reeves et al. (1993) for the YBCO compound. The main challenge was to evaluate the electronic contribution to the heat capacity at high temperatures where the heat capacity is dominated by the lattice. The lattice contribution was calculated using the phonon density of states obtained by neutron scattering. As a result, the value of $\lambda > 2.5$ was obtained, which means that there is strong electron-lattice coupling sufficient to provide high $T_c$.

## IV. **Polaronic effect**

### A. Polarons and isotope effects

A strong electron-lattice interaction could lead to specific polaronic effects. The concept of polarons was introduced and studied by Pekar (1946) and Pekar and Landau (1948). A polaron can be created if an electron is added to the crystal with a small carrier concentration (see,e.g., Ashcroft and Mermin, 1976). Because of strong local electron-ion interactions, the electron appears to be trapped and can be viewed as being dressed in a "heavy" ionic "coat". In reality, we are dealing with a strong (nonlinear) manifestation of the electron-lattice interaction.

The concept of polarons is an essential ingredient of the physics of high $T_c$ oxides. In fact, the formation of a Jahn -Teller polaronic state was a main motivation for Bednorz and Mueller to search for superconductivity in these systems, and this led to their breakthrough discovery. They gave a significant amount of credit to the paper on Jahn-Teller polarons by Hock et al. (1983).

The formation of polaronic states is a strong nonadiabatic phenomenon. As we know, the usual adiabatic method [Born-Oppenheimer approximation (1927); see also reviews by Born and Huang (1954) ; Bersuker (1984); Kresin et al.(1993)] allows us to separate electronic and ionic motions. Indeed, this approximation is based on the fact that in metals the ionic motion is much slower than the motion of electrons (the inequality $\tilde{\Omega}/E_F << 1$ is a condition of applicability of this approximation), and it allows us as a first step, to neglect the kinetic energy of the ions and to study the electronic structure for a "frozen" lattice. The electronic energy (electronic terms) appears to be a function of the ionic positions [ $\varepsilon_{el} = \varepsilon_n(\vec{R})$]. Next, one can study the ionic dynamics; it turns out that the electronic terms $\varepsilon_n(\vec{R})$ form the potential for the ionic motion. The total wave function $\Psi$ can be written as a product : $\Psi = \psi_{el}\phi_{ionic}$. However, such a separation of electronic and ionic terms is impossible for polaronic states. Speaking of the high Tc cuprates, it is important that oxygen ions actively participate in the formation of such states. Note that these ions play a unique role in the lattice dynamics, because they are the lightest elements in the cuprates. Polaronic effect increases the phase space for pairing virtual transitions (Kresin, 2009)

The implications for the isotope effect because of the presence of polaronic (bound electron-ionic) states was described by Kresin and Wolf (1994a)

One can assume that an oxygen ion is characterized not by the usual local minimum of the potential, but rather by two closely spaced minima (Fig. 5; double-well structure). Note that the "double-well" structure is a characteristic feature for both the in-plane and apical ions (see Fig. 1). Such a double-well structure has been observed experimentally using the x-ray absorption fine- structure technique (Haskel et al., 1997); see Fig.6.

Note that the double-well structure is a result of the crossing of electronic terms. The ionic configuration at this crossing corresponds to a degeneracy of the electronic states which is a key ingredient of the Jahn-Teller effect (see, e.g., Landau and Lifschitz,1977).

We start with an apical oxygen. The dynamics of the apical oxygen ions plays an essential role in these compounds [see, e.g., Mueller (1990)]. The cuprates are doped materials, and because of it charge transfer through this ion is an important factor. One can show (Kresin and Wolf,1994a,1994b) that the doping and therefore, the carrier concentration are affected by an isotopic substitution. Since the value of $T_c$ depends strongly on carrier concentration [$T_c \equiv T_c(n)$], we are dealing with a peculiar isotopic dependence of $T_c$. If the charge transfer occurs in the framework of the usual adiabatic picture, so that only the carrier motion is involved, then the isotope substitution does not affect the forces and therefore does not change the charge- transfer dynamics.  However, strong nonadiabaticity changes the picture rather dramatically. The electronic and nuclear motions are not separable, and in this case the charge transfer is a more complex phenomenon that does involve nuclear motion.

The presence of two close minima means that the degree of freedom describing the ionic motion corresponds to electronic terms crossing (see Fig. 5). The charge transfer in this case is described by polaronic motion, that is, by the motion of the nearly bound electron-ionic unit (this can be described as a dynamic polaron). Note that a similar effect leads to the isotope effect in manganites (Gor'kov and Kresin, 2004).

Qualitatively, the charge transfer for such nonadiabaticity can be visualized as a multistep process: first the carrier makes a transition from the chain site to the apical oxygen, then the apical oxygen transfers to another term (see Fig. 5), and this is finally followed by the transition of the carrier to the plane. The second step is affected by the isotope substitution. For the entire crystal ,it can be viewed as the motion of a polaron (dynamic polaron).

In order to describe this phenomenon, it is convenient to use a so-called "diabatic" representation (see, e.g., O'Malley ,1967; Kresin and Lester,1984; Dateo et al.,1987). In this representation , we are dealing directly with the crossing of electronic terms. The operator $\hat{H}_{el.} = \hat{T}_{\vec{r}} + V(\vec{r}, \vec{R})$ [ $\hat{T}_{\vec{r}}$ is a kinetic energy operator, $V(\vec{r}, \vec{R})$ is a total potential energy,and $\vec{r}$ and $\vec{R}$ are the electronic and nuclear coordinates, respectively] has nondiagonal terms (unlike the usual adiabatic picture when $\hat{H}_{el}$ is diagonal). The charge transfer in this picture is accompanied by the transition to another electronic term. Such a process is analogous to the Landau-Zener effect (see Landau and Lifshitz, 1977).

The total wave function can be written in the form

$$\Psi(\vec{r}, \vec{R}, t) = a(t)\ \Psi_1(\vec{r}, \vec{R}) + b(t)\ \Psi_2(\vec{r}, \vec{R}) \qquad (4.1\ )$$

Here

$$\Psi_i(\vec{r}, \vec{R}) = \psi_i\ (\vec{r}, \vec{R})\ \Phi_i\ (\vec{R})\ ,\ i = \{1,\ 2\}$$

$\psi_i(\vec{r}, \vec{R})$, $\Phi_i(R)$ are the electronic and vibrational wave functions that correspond to two different electronic terms (see Fig. 5).

In the diabatic representation, the transition between terms isdescribed by the matrix element $V_{12}$, where $\hat{V} = \hat{H}_r$. One can show that

$$V_{12} \cong L_0\ F_{12} \qquad (4.2)$$

where $L_o = \int d\vec{r}\ \psi_2^*(\vec{r},\vec{R})\ \hat{H}_r\ \psi_1(\vec{r},\ \vec{R})\ \big|_{R_o}$ is the electronic constant

($R_0$ correspond to the crossing configuration), and $F_{12} = \int \varphi_2^*\left(\vec{R}\right)\varphi_1\left(\vec{R}\right)d\vec{R}$ is the so-called Franck-Condon factor. The presence of the Franck-Condon factor is a key ingredient of our analysis. Its value depends strongly on the ionic mass and, therefore is affected by the isotope substitution. The calculation (Kresin and Wolf,1994a) leads to the following expression for the isotope coefficient:

$$\alpha = \gamma\ \frac{n}{T_c}\ \frac{\partial T_c}{\partial n}\ , \qquad (4.3)$$

where $\gamma$ has a weak logarithmic dependence on ionic mass M. Therefore, the polaronic isotope effect ($\alpha \equiv \alpha_{ac}$; $\alpha_{ac}$ corresponds to the apical oxygen ion) is determined by the dependence of $T_c$ on n, where n is the carrier concentration. A strong nonadiabaticity (the apical oxygen in YBCO is in such a nonadiabatic state) results in a peculiar polaronic isotope effect.

The impact of the isotope substitution $O_{16} \rightarrow O_{18}$ on the in-plane oxygen ($\alpha \equiv \alpha_p$) looks different. The corresponding vibrational mode is directly affected by the isotopic substitution and thus makes a direct contribution to the pairing as in the normal isotope effect. In addition, the polaronic nature of the carriers in the planes also provides a novel isotope effect due to an increase ($O_{16} \rightarrow O_{18}$) in the carriers effective mass, which leads to a change in the value of $T_c$. Therefore, the polaronic effects are essential for both the in-plane and apical oxygen sites (Bussmann-Holder and Keller,2005). According to Eq. (4.6), at optimum doping $\partial T_c / \partial n = 0$ and, therefore, the apical oxygen ion does not make any contribution. In this case, the main contribution comes from the in-plane oxygen. This was confirmed by site-selected experiment (Zech et al.,1994). One can expect that the value of $\alpha_{ac}$ increases for the region that is far from optimum $T_c$. It is important (see, e.g., review by Keller ,2005) for such experiments that the isotope effect should be measured on the same sample to guarantee that the doping level (oxygen concentration) is unchanged with the isotopic substitution.

Note that there is no one-to-one correspondence between the amount of oxygen and the in-plane carrier concentration. The in-plane carrier concentration can be affected by the isotopic substitution on the apical site. Because of the polaronic effect, the probability of tunneling becomes different and this leads to the redistribution of the total electronic wave function between the Cu-O plane and the charge reservoir.

The site-selected experiments have been performed for $Y_{1-x}Pr_xBa_2Cu_3O_{7-\delta}$ samples ( Khasanov et al., 2003; Keller,2003). The Pr substitution leads to a depression in $T_c$; the samples studied have $T_c \approx 44$ K. Both $\alpha_p$ and $\alpha_{ac}$ are large relative to their values at x=0; the in-plane term $\alpha_p$ is larger than $\alpha_{ac}$. This increase can be caused by mixed valence of the Pr ions as well as a pairbreaking effect caused by magnetic moments on the Pr site. Indeed, pairbreaking affects the value of the isotope coefficient ( Carbotte,1991; Kresin et al.,1997). In connection with this, it would be interesting to carry out the site-selective experiments for samples with different oxygen contents.

The polaronic effect also leads to the possibility of observing an unusual isotopic dependence of the penetration depth, since this quantity also depends on the carrier concentration as well as on the effective mass. This effect was introduced theoretically by Kresin and Wolf (1994b) [ see also Bill et al. (1998) ] and observed experimentally by Zech et al. (1996) and Khasakov et al.( 2004);Keller,(2008). The muon-spin rotation technique (μSR;see review by Keller, 1989) was employed; this method allows the direct determination of the penetration depth. According to recent experimental data  (Khasanov et al.,2006; Keller, 2006) , the correlation between isotope effects on $T_c$ and penetration depth can be explained by the interplay of both polaronic channels affecting the carrier concentration and effective mass. It is clear that these data demonstrate the importance of polaronic effects.

Another polaronic effect that also reflects the importance of the electron-lattice interaction was observed by Oyanagi (2007) using x-ray absorption spectroscopy . This method (see Bianconi et al.,1996; Oyanagi et al.,2007) reveals that doping leads to displacement of oxygen atoms, and this demonstrates the impact of the electron-phonon interaction. More specifically, Oyanagi (2007) measured the Cu-O radial distribution function. Upon cooling, a sharp decrease in this function at $T_c$ was observed. Such a sharpening in the radial distribution function reflects the appearance of correlated motion of oxygen ions and is connected with the phase coherence of the electronic subsystem. Such a large impact of the pairing on the dynamics of the ions is caused by the fact that it is impossible to separate the electronic and ionic degrees of freedom, and again this corresponds to the propagation of a dynamic polaron.

## B. "Local" pairs: bi-polarons, U-centers and the BEC-BCS scheme

A bipolaron represents a local structure that can be viewed as a bound state of two polarons. This type of structure is supposedly caused by a very strong electron-lattice interaction. Therefore, the bipolaronic scenario represents an extreme case of electron-phonon (lattice) dynamics. It is interesting to note that a scenario of "local" pairs was proposed as an explanation of superconductivity even before the BCS theory (Schafroth,1955). A more rigorous concept of a bipolaron , which is a bound state of two polarons ,was introduced by Vinetskii (1961) and Eagles (1967). The qualitative picture of bipolaronic superconductivity is rather elegant and is very different from the conventional BCS concept. The main difference is the nature of the normal state. As we know, the starting point of the BCS picture is that in the normal state (above $T_c$ or above the critical field) we are dealing with the usual fermions (delocalized electrons) and , correspondingly, with a Fermi surface. According to the bipolaronic picture, the normal state represents a Bose system formed by pairs of polarons: pairing occurs in real space. As a result, the nature of the phase transition at $T_c$ is entirely different. According to the bipolaronic scenario, we are dealing with the Bose-Einstein condensation of bosons , whereas the formation of pairs (Cooper pairs) in usual superconductors occurs at $T_c$. The Cooper pair is formed by two electrons with opposite momenta, so that the pairs are formed in momentum, not real space.

A more detailed model of bipolaronic superconductivity, namely the picture that the bosons (bipolarons) formed on a lattice could form a superconducting system , was proposed by Alexandrov and Ranninger (1981). A small value of the coherence length, along with a low carrier concentration, typical in the cuprates made the bipolaronic picture attractive. And, indeed, after the discovery of high $T_c$ cuprates several (see, e.g., Emin,1989; Broyles et al.,1990; Micnas et .al.,1990; Alexandrov and Mott ,1994; and Alexandrov and Andreev, 2001) proposed such a picture and developed many of its aspects. However, Chakraverty et al. (1998) later came to the conclusion that this scenario is not applicable to the cuprates, because of its incompatibility with experiments. The value of the effective mass that is required for the observed critical temperature appears to be drastically different from the observed one. Moreover, the bipolaronic picture requires a bosonic nature of the carriers. This factor is even more important, since it contradicts the existence of the Fermi surface that was established experimentally (Marshall et al.,1996; Ding et al.,1996). Note that at present the evidence for the existence of a Fermi surface is even stronger (see, e.g., Hussey et al., 2003).

Note also that the statement about the lattice instability leading to the formation of bipolarons at $\lambda \gtrsim 1$ was based on the usual Froelich Hamiltonian (see Sec.II). According to rigorous adiabatic theory (Geilikman,1975), this approach is valid only if $E_F << \bar{\Omega}$, which is not the case for conventional superconductors, and is also not the case for the cuprates where $E_F \sim 1$ eV, $\bar{\Omega} \sim$ 10-50 meV.

A more general picture was described by Mueller et al. (1998), see the review by Mueller (2007). According to this approach, the high $T_c$ compound contains two components: bipolarons and free fermions. The presence of free fermions explains the presence of the Fermi surface. As a whole, the model describes many experimental results.

A picture of negative U centers formed by two electrons localized on the same lattice site was introduced by Anderson (1975) to study amorphous semiconductors. The appearance of such centers is caused by a strong local electron-lattice interaction. After the discovery of the high $T_c$ cuprates it was suggested (Schuttler et al.,1987) that the presence of Uimpurities can result in a large increase of $T_c$. The theoretical study by Oganesyan et al. (2002) demonstrated that the U centers can provide the resonant tunneling channel between the $CuO_2$ layers (see Geballe ,2006).

Another interesting approach is concerned with a scenario that is intermediate between the Bose-Einstein condensation (BEC) and Cooper pairing (BCS). Such a generalization was considered initially by Leggett (1980) and later by Nozieres and Schmitt-Rink (1985) and Nozieres (1995) . The properties of a Fermi gas with an attractive potential have been studied as a function of the coupling strength. BEC and BCS cases correspond to two limits (strong and weak coupling). It is remarkable that the evolution between these two limits is smooth. Nozieres and Schmitt-Rink (1985) studied not only the evolution of the ground state, but also the change in the transition temperature, and theystressed the importance of individual excitations for the Cooper pairing channel versu collective excitations for the BEC case.

All the examples described in this section are directly related to the impact of the lattice and the electron-lattice interaction on the electronic subsystem and its superconducting state. The possibility of the appearance of local pairs and the impact of such factors as the presence of two components or U centers in the cuprates or other complex systems deserve additional theoretical and, especially, experimental study.

## V. Phonon-plasmon mechanism

In this section, we discuss the phonon mechanism which is combined with a peculiar plasmon contribution. Plasmons represent collective electronic modes; they can be visualized as collective electronic oscillations with respect to positive ionic background (see, e.g., Ashcroft and Mermin,1976). For simple metals the value of the plasmon frequency is rather high ($\approx$ 5-10 eV), and it depends weakly on momentum. Metals with a complex band structure display additional low-lying plasmon branches. The layered conductors also have a peculiar structure of their plasmon spectrum, and in this section we focus on this case.

The plasmon mechanism implies that pairing occurs via the exchange of plasmons; in other words, plasmons play a  role similar to that of phonons. Here we discuss the situation when pairing is provided by contributions of both channels, that is, by phonons and  plasmons.

 The plasmon mechanism of superconductivity has been studied previosly ( Froelich, 1968; Geilikman, 1966, Ihm et al., 1981). The interelectron coupling is provided by the acoustic plasmon branch; this mode corresponds to the collective motion of the light carriers with respect to the heavy ones (e.g., for the case of two overlapping different bands). For the cuprates such a channel was studied by Ruvalds (1987). Another possibility was studied by Ashkenazi et al. ( 1987). It has been proposed that a charge density-wave instability will lead to softening of the plasmon branch, and this leads to strong pairing.

Here we focus on the plasmon spectrum specific for layered conductors. This question is interesting not only for the study of the cuprates. Indeed, the past few years have witnessed the discovery of many new superconducting materials: high temperature cuprates, fullerides, borocarbides, ruthenates, $MgB_2$, metal-intercalated halide nitrides, intercalated $Na_xCoO_2$, etc. Systems such as organics, heavy fermions, and nanoparticles have also been studied intensively. Many novel systems belong to the family of layered (quasi-two-dimensional) conductors and are characterized by strongly anisotropic transport properties. One can raise an interesting question: why is layering a favorable factor for superconductivity? One can show (Kresin,1987b; Kresin and Morawitz,1988; Bill et al., 2003) that layering leads to a peculiar dynamic screening of the Coulomb interaction. Layered conductors have a plasmon spectrum that differs fundamentally from three-dimensional metals. In addition to a high-energy ''optical'' collective mode, the spectrum also contains an important low -frequency part   ("electronic" sound, see  Fig. 7), see Fetter (1974), Kresin and Morawitz (1990), Morawitz et al. (1993). The screening of the Coulomb interaction is incomplete and the *dynamic* nature of the Coulomb interaction becomes important. The contribution of the plasmons *in conjunction* with the phonon mechanism may lead to high values of $T_c$.

We consider a layered system consisting of a stack of conducting sheets along the *z* axis separated by dielectric spacers. Because of the large anisotropy of the conductivity, it is a good approximation to neglect transport between the layers. On the other hand, the Coulomb interaction between charge carriers is effective both *within* and *between* the sheets. In order to calculate the superconducting critical temperature $T_c$, one can use the equations for the superconducting order parameter [cf.Eq.(2.6)]:

$$\phi_n\left(\vec{k}\right) = T \sum_{m=-\infty}^{\infty} \int \frac{d\vec{k}}{(2\pi)^3} \Gamma\left(\vec{k},\vec{k}';\omega_n - \omega_m\right) \frac{\phi_m\left(\vec{k}'\right)}{\omega_m^2(\vec{k}') + \xi_{\vec{k}'}^2}\big|_{T_c} \qquad (5.1)$$

Here $\phi_n\left(\vec{k}\right) = \Delta_n\left(\vec{k}\right)Z, \omega_n(\vec{k}) = \omega_n Z, \Delta_n \equiv \Delta(\omega_n)$. We shall not write out the expression for Z. The interaction kernel $\Gamma$ can be written as a sum of electron-phonon and Coulomb interactions. The Coulomb term contains the plasmon excitations and the usual static repulsion.

A detailed analysis (Bill et al., 2003) based on Eq. (5.1) shows that the impact of dynamic screening is different for various layered systems. For example, for the metal -intercalated halide nitrides (see, e.g., Yamanaka et al.,1998) the plasmon contribution dominates. As for the cuprates, the plasmon contribution is not so crucial but is noticeable: about 20% of the observed value of $T_c$ is due to acoustic plasmons. The main role is played by phonons ,and their impact leads to high value of $T_c$.

## VI. **Electron-phonon interaction and the "pseudogap" state**.

A study of the "pseudogap" state of the high $T_c$ cuprates has attracted much interest. This issue is very interesting and is still controversial. As we know, the superconducting state of usual superconductors is characterized by zero resistance, anomalous diamagnetism which strongly depends on temperature, by an energy gap, etc. These features are absent above $T_c$ , in the normal state [ except for the effect of  fluctuations near $T_c$; see, e.g., Larkin and Varlamov, 2005] . The situation for the cuprates, especially in the underdoped state, is entirely different. According to many experimental results, one can observe, above $T_c$, along with normal resistance such properties as an energy gap, anomalous diamagnetism, isotopic dependence of the "pseudogap" temperature $T_c^*$, a "giant" Josephson effect, etc., that is, many features that are characteristic of a superconducting state.

It is important to realize that, because of doping (carriers are added by substitution or nonstoichiometry), we are dealing with an intrinsically inhomogeneous system. As a result, the compounds display phase separation (Gor'kov and Sokol, 1987) [ see also, e.g., Sigmund and Mueller (1994)], that is, the coexistence of metallic and insulating phases. According to our approach (Ovchinnikov et al., 1999; see also review by Kresin et al.,2006), upon cooling below some characteristic temperature $T_c^*$, the metallic phase becomes inhomogeneous and represents a mixture of superconducting and normal regions. As temperature decreases toward $T_c$, the size of the superconducting regions and their number increase. At $T=T_c$ one can observe the percolative transition, that is, the formation of macroscopic superconducting regions. Such a picture was directly observed by Igushi et al. (2001) using the STM technique with magnetic imaging.

Recently, Gomes et al. (2007), using a specially designed variable temperature STM, observed that pairing occurs initially in small regions and can persist at temperatures that greatly exceed the resistive $T_c$ (for $Bi_2Sr_2CaCu_2O_{8+\delta}$, the superconducting nanoregions were observed at $T \approx 160$ K). The observation confirms our predictions. Recent bulk $\mu$SR data (Sonier et al.,2008) also support our picture.

One might think that the inhomogeneous nature of the cuprates is an important feature, but it is not directly relevant to the pairing mechanism. However, recent experiments by Gomes et al. (2007) appear to be important also from this point of view. They measured local values of $T_c$ and the gap; it has been observed that the ratio $2\Delta/T_c$ is rather large [ $2\Delta/T_c \approx 8$; here $\Delta \equiv \Delta(0)$ is the energy gap at T=0K ]. Such a large value corresponds to the strong-coupling case and is consistent with the electron-phonon scenario for pairing. Indeed, the ratio $2\Delta/T_c$ is directly related to the strength of the interaction. According to the BCS theory (weak coupling; $\lambda \ll 1$), this ratio is universal and is given by $2\Delta/T_c$=3.52. An increase in $\lambda$ leads to an increase in this ratio . For example, for Pb ($\lambda \approx 1.5$) $2\Delta/T_c \approx 4.3$, and for $Pb_{0.7}Bi_{0.3}$ ($\lambda \approx 2$) $2\Delta/T_c \approx 4.85$ [see the review by Wolf (1985)]. The ratio can be calculated using the general equation [ Geilikman and Kresin,1966;see the reviewsGeilikman et al.,1975; Carbotte,1990, Kresin and Wolf,1990]:

$$\frac{2\Delta}{T_c} = 3.52\left[1 + 5.3\left(T_c/\tilde{\Omega}\right)^2 \ln\left(\tilde{\Omega}/T_c\right)\right]$$ (6.1)

According to Kresin (1987c), the ratio $2\Delta/T_c$ for strong coupling lies above the BCS value $(2\Delta/T_c)_{BCS}$=3.52 and below the upper limit $(2\Delta/T_c)_{max.}$= 13.4. The measured value of $2\Delta/T_c$=8 corresponds to strong electron-phonon coupling with values of $\lambda \approx 3\text{-}3.5$ which are quite large and are sufficient to explain the observed value of $T_c$. It is interesting to note that this value was observed in a system that contains nan-regions (see below,Sec..VIII).

A large isotope effect on $T_c^*$ ($T_c^* \propto M^{-\alpha}$; $\alpha \approx -2.2 \pm 0.6$) has also been observed (Lanzara et al.,1999; Furrer, 2005 ).This can be explained by the presence of superconducting regions ( Kresin et al., 2006) and by the polaronic effect (see Sec.IV) , and can be described by a relation similar to (4.3), that is,

$$\alpha = \gamma \frac{n}{T_c^*} \frac{\partial T_c^*}{\partial n} \qquad (6.2)$$

The experimental observation of the isotope effect on $T_c^*$ also reflects the fact that superconducting pairing persists above the resistive transition. It is interesting to note that the experimentally measured isotope coefficient has a negative sign. This can be explained by Eq.(6.2) and by the fact that an increase in doping in the underdoped region leads to a decrease in the value of $T_C^*$ (at optimum doping $T_C^* \cong T_C$); as a result, $\alpha < 0$.

VII. **Proposed experiment**

All experiments described above (Secs.III,VI) are interesting and informative and provide strong evidence for the contribution of the electron - phonon interaction to the superconductivity for many of the newly discovered superconductors, especially the cuprates. However, one can propose a different experiment (Ovchinnikov and Kresin ,1998; Ovchinnikov et al., 1998) that will allow the unambiguous determination of the coupling boson (excitation) in the cuprate superconductors. This method is based on the generation and detection of the appropriate boson and is analogous to the experiments on the generation of phonons by conventional BCS superconductors.

The method is based on the technique of using Josephson junctions for the generation of phonons (Eisenmerger and Dayem, 1967; Eisenmerger 1969; Dynes et al., 1971; Dynes and Narayanamurti,1973). One can modify this technique for any boson contributing to the pairing. The generation of excitations caused by pair recombination can be used as a signature of the mechanism of pairing. A nonequilibrium superconducting state is formed by incoming radiation. The creation of excited quasiparticles is followed by a relaxation process. By the end of this process , a noticeable number of quasiparticles are concentrated at or very near the energy gap edge, $\varepsilon \approx \Delta$, where $\Delta$ is the pairing gap. The final stage of relaxation is the recombination of Cooper pairs. For conventional superconductors , this stage is accompanied by radiation of phonons.

In a classic experiment (Eisenmenger and Dayem, 1967; Eisenmenger, 1969)  the generation and detection of phonons propagating through a sapphire substrate was demonstrated  using two Josephson junctions located diametrically on opposite sides of a cylindrical sapphire block. This pioneering work was followed up by several investigations thatdeveloped an understanding of the details of the spectroscopy of the phonons generated and detected by similar means. The time and energy distribution of the phonons that were emitted were studied by such experiments. The study was aimed at the generation of almost monochromatic phonons. We now to look at such experiments from a different point of view. Indeed, these experiments were possible only because phonons were responsible for pairing in the electrodes of the emitting junction and are thus emitted when quasiparticle excitations relax to the gap edge and recombine to form pairs. In other words, one can observe the recombination of electrons with energies near the gap edge; these electrons can form Cooper pairs and this process is accompanied by radiation of phonons with $\hbar\omega \approx 2\Delta$.

One can raise the following question: why are other excitations not radiated, only phonons? The answer is obvious and directly reflects the fact that phonons form the glue for pairing. In fact, radiation of phonons created by recombination is an additional support for the phonon mechanism of pairing in conventional superconductors. If pairing is provided, e.g., by magnetic excitations, the recombination would be accompanied by radiation of magnons.

One can propose a series of experiments analogous to these pioneering efforts; such experiments can provide an unambiguous determination of the appropriate boson responsible for superconductivity in the cuprates. It is crucial that the proposed experiments can distinguish between phonon and nonphonon (e.g., magnon) coupling based on the selection of the propagation medium.

Assume for the moment that superconductivity in the cuprates is mediated by phonons. Then we propose the following experiment. On one side of a high quality sapphire (or other nearly defect-free single crystal substrate) one can prepare a Nb or NbN tunnel junction as a detector of phonons. This detector will be most sensitive to phonons that are above the gap energy $2\Delta$ of the electrodes. Phonons with energy lower that the appropriate gap energy will be filtered out since they will not break pairs and will not be detected. On the other side of the substrate, we prepare a cuprate junction or weak link that can be biased into the normal state by a current or infrared pulse. After the current or light has been removed, the quasiparticles generated will relax very rapidly to the gap edge and as they recombine to form pairs they will emit $2\Delta$ (T,**k**) phonons. The gap may be anisotropic so that the phonon energy will be dependent on where in **k** space the quasiparticles are located. In addition, many cuprates appear to have a gapless superconducting density of states. Although  this density of states is peaked near some value of $\Delta$ , it is characterized by the presence of  states all the way down to E=0.  In any event, a large number of phonons will have energies well above the gap of the detector , which is a conventional, low $T_C$ superconductor. In this case,  we expect to see a very well defined signal similar to what was observed for conventional junctions. To calibrate the experiment, we propose that on the very same substrate (prior to the deposition of the cuprate junction) we prepare a Nb or NbN junction as the emitter and perform a replication of the original Eisenmenger and Dayem experiment to estimate the sensitivity.  Thus the observation of a signal from the cuprate, similar in magnitude and temporal behavior to that of the control junction, would be extremely strong evidence that phonons were the primary excitation from the recombination of excited quasiparticles.

The relaxation process of excited quasiparticles consists of several stages. As a result of electron-electron (first stage) and electron-phonon (second stage) collisions, a number of quasiparticles near the value E=2Δ will appear. One can show (Ovchinnikov and Kresin,1998 ) that this process is described by

$$(\partial W / \partial t) = -\lambda (\Delta / \Omega)^2 W^2 \qquad (7.1)$$

Here $W$ is the number of quasiparticles, $\lambda$ is the electron-phonon coupling constant, and $\Omega \approx \Omega_D$. Using Eq. (7.1) , it is easy to determine the function $W$ (t), and then the number of generated phonons ΔN,

$$\Delta N(t) = aW(0)t[1+at]^{-1}$$
$$a \equiv \lambda (\Delta / \Omega)^2 W(0) \qquad (7.2)$$

Here $W$ (0) = $W$ (t=0); t=0 corresponds to the beginning of the second stage.

A significantly smaller signal would indicate that phonons were not the primary recombination excitation but might be secondarily produced by the decay of the primary boson. In this case ,the energy of the secondary phonons will be much smaller than the gap in the cuprate junction and also smaller than the gap in the NbN , which would mean they would not be detected. Such a small signal would indicate that the pairing boson might be a spin fluctuation or magnon , which then could be confirmed by another series of experiments.

If we now assume that spin fluctuations are the primary pairing excitation, then we would replace the substrate that was a very good phonon propagator with a substrate that would not support the propagation of high-energy phonons but was magnetic and would be an excellent propagator of magnons. Perhaps single -crystal yttrium iron garnet (YIG) with appropriate impurities could be prepared into such a substrate. In fact, cuprate films have already been prepared on such YIG substrates. The same NbN junction would be placed on one side of this substrate and the same two experiments should be performed. The conventional emitter should give a very small signal, whereas the cuprate signal should be much larger, an indicator that magnons are the primary recombination excitation.

Note also that according to many experiments the energy range for phonons and magnons is similar. For example, for YBCO both the phonon and magnon spectra range from E=0 up to E=40-50 meV. Therefore, both channels are available for the relaxation process, and the dominance of one of them means that the electrons mainly interact with bosons (phonons or magnons) corresponding to this channel. In addition, since $2\Delta \approx E_{ph.} \approx E_{magn.}$, the pairing (interaction with virtual excitations) and the relaxation are governed by similar matrix elements.

A sharp signal will be observed if the superconducting oxides have a well defined energy gap. From this point of view, the Nd-based cuprate and BaKBiO systems could be selected for the initial study. In accordance with Murakami et al. (1994), the LaSrCuO compound also has a sharp gap and could represent a good candidate for this experiment as well. As for YBCO and other cuprates, they are usually characterized by a gapless spectrum. Nevertheless, the signal generated by the recombination can be detected. In addition, as noted above, for these materials $2\Delta \approx E_{ph}$, and the appearance of phonons with a frequency similar to that for the virtual transitions is a strong indication of a key contribution of the phonon mechanism.

VIII. **Superconducting State of Nanoclusters**

  A. Nanoparticles and size quantization

As noted above (Sec.II), the electron-lattice interaction, in principle, can provide a high value of the critical temperature. This aspect of the interaction is apparent in various superconducting systems. For example, $MgB_2$ , which has a relatively high $T_c \approx$ 42 K, is generally accepted to be a phonon mediated superconductor.

As we know from our understanding of conventional superconductors, an increase in $T_c$ can be achieved by an increase in the density of states at the Fermi level; this is natural, since the density of states enters as a factor in the expression for the coupling constant [see Eq. (2.7)]. Historically, the highest value of $T_c$ for conventional superconductors was observed in A-15 compounds. These large values are caused by the presence of a Van Hove singularity in the density of states (DOS), that is, by a sharp peak in the DOS at the Fermi level (Labbe et al.,1967).

It has also been observed (Kuhareva,1962; Strongin et al., 1965) that the $T_c$ of Al films (~ 2.1 K) can be nearly double the value for bulk samples. Even larger increases ($T_c \approx$ 3 K) were observed for granular Al (Deutcher et al., 1973). These increases can be explained by size quantization and corresponding increase in the effective density of states in films and isolated granules; this was explained by Kresin and Tavger (1966) for films and by Parmenter (1968) for granular structures.

The most distinctive feature of nanoparticles is the discrete nature of their electronic spectra. The superconducting state of nanoparticles has been studied by Tinkham et al. (1995); see review by von Delft and Ralph (2001). They studied nanoparticles that were placed inside a tunneling barrier that contained $N \approx 10^4$-$10^5$ delocalized electrons.

We focus below on smaller nanoparticles, so-called nanoclusters, with $N \approx 10^2$-$10^3$ delocalized electrons.

B. Nanoclusters and the high $T_c$ state

High $T_c$, potentially up to room temperature, should be observed for specific metallic nanoclusters (Ovchinnikov and Kresin, 2005 ; Kresin and Ovchinnikov, 2006). This attractive possibility could be realized thanks to a remarkable feature of metallic clusters, namely the shell structure of their electronic spectra. This phenomenon was discovered by Knight and collaborators (1984).  Initially the presence of a shell structure was observed for alkali-metal clusters. Later  the presence of energy shells has been detected for many other nanoclusters including   Al, Ga, Zn, Cd, and In [see the review by de Heer  (1993)]. The importance of the shell structure for superconductivity was discussed by Knight (1987), and Friedel (1992) , who stressed the possibility of a large increase in $T_c$. The appearance of a superconducting state requires that $\delta E \lesssim \Delta$ (Anderson, 1959), where $\Delta$ is the gap parameter and $\delta E$ is the spacing between discrete electronic levels. One important aspect of the shell structure is that this criterion could be met.

As noted above, metallic clusters contain delocalized electrons whose states organize into shells, similar to those in atoms or nuclei [ see, e.g., review by Frauendorf and Guet (2001)]. In some clusters, shells are completely filled all the way up to the highest occupied shell (HOS), e.g., those with $N=N_m=20, 40, 58, 92, 138, 168,....$ These values are known as "magic" numbers. Such clusters are spherical. The electronic states in such "magic" clusters are labeled by their orbital momentum $l$ and radial quantum number $n$. Cooper pairs are formed by electrons with opposite projections of orbital momentum [such a pairing is similar to that in atomic nuclei, see, e.g., review by Ring and Schuck (1980)]. If the orbital momentum $l$ is large, the shell is highly degenerate ($2(2l+1)$ is large). This factor drastically increases the effective density of states. In addition, the energy spacing $\Delta E$ between neighboring shells varies, and some of them are separated by only a small $\Delta E$. One can show that the combination of high degeneracy and a small energy spacing between the highest occupied shell (HOS) and the lowest unoccupied shell (LUS) leads to the possibility of a large increase in the strength of the superconducting pairing interaction in the corresponding clusters. Qualitatively, this can be understood in the following way. If the HOS is highly degenerate, this means that the shell contains many electrons, which can be viewed as a sharp peak in the density of states at the Fermi level. An increase in the density of states leads to an increase in the value of the electron-phonon coupling constant; this can be seen directly from Eq. (2.17). As a result, one can obtain very large values of $T_C$. This situation is similar to that studied by Labbe et al. (1967) for bulk materials; the presence of a peak in the density of states results in a noticeable increase in $T_C$.

The equation for the pairing order parameter $\Delta(\omega_n)$ has the following form [cf. Eq.(2.6)]:

$$\Delta(\omega_n)Z = \eta \frac{T}{2V} \sum_{\omega_{n'}} \sum_s D(\omega_n - \omega_{n'}) F_s^+(\omega_{n'})$$  (8.1)

$D(\omega_n - \omega_{n'}, \tilde{\Omega}) = \tilde{\Omega}^2 \left[ (\omega_n - \omega_{n'})^2 + \tilde{\Omega}^2 \right]^{-1}$    and

$F_s^+(\omega_{n'}) = \Delta(\omega_{n'}) \left[ \omega_{n'}^2 + \xi_s^2 + \Delta^2(\omega_{n'}) \right]^1$    are the vibrational propagator and the Gor'kov pairing function (1958), respectively, $\xi_s = E_s - \mu$ is the energy of the sth electronic state referred to the chemical potential $\mu$, V is the cluster volume, $\eta = \langle I^2 \rangle / M\tilde{\Omega}^2$ is the so-called Hopfield parameter [cf.Eq. (2.7)], and Z is the renormalization function.

Eq. (8.1) contains a summation over all discrete electronic states. For "magic" clusters which have a spherical shape, one can replace the summation over states by summation over the shells: $\sum_s \rightarrow \sum_j G_j$, where $G_j$ is the shell degeneracy: $G_j=2(2l_j+1)$, where $l_j$ is the orbital momentum. Then Eq. (8.1) can be written in the form

$$\Delta(\omega_n)Z = \lambda \frac{2E_F}{3N} \sum_{\omega_{n'}} \sum_j G_j \frac{\tilde{\Omega}^2}{\tilde{\Omega}^2 + (\omega_n - \omega_{n'})^2} \frac{\Delta^2(\omega_n)}{\omega_n^2 + \xi_j^2} |T_c \qquad (8.2)$$

We used the expression for the bulk coupling constant $\lambda = \nu \langle I^2 \rangle / M\tilde{\Omega}^2$ [Eq. (2.7)] , where $E_F$ is the Fermi energy. Note that the characteristic vibrational frequency is close to the bulk value because pairing is mediated mainly by the short-wave part of the vibrational spectrum.

If the shell is incomplete, the cluster undergoes a Jahn-Teller deformation, so that its shape becomes ellipsoidal, and the states s are classified by their projection of the orbital momentum $|m| \leq l$, and each level contains up to four electrons (for $|m| \geq 1$ ). Note that in the weak coupling case ($\eta$/V <<1 and correspondingly $\pi T_c << \tilde{\Omega}$ ), one should put in Eq. (7.1) Z=1, D=1, recovering the usual BCS scenario.

Equation (8.1) looks similar to the equation appearing in the theory of strong coupling superconductivity, see Eq.(2.6), but is different in two key aspects. First, it contains a summation over discrete energy levels $E_S$ whereas for a bulk superconductor one integrates over a continuous energy spectrum (over $\xi$). Second, as opposed to a bulk superconductor, we are dealing with a finite Fermi system, so that the number of electrons N is fixed. As a result, the position of the chemical potential $\mu$ differs from the Fermi level $E_F$ and is determined by the values of N and T.

It is essential that the value of the critical temperature $T_c$ is determined by parameters that can be measured experimentally. These parameters as follows: the number of valence electrons N, and the energy spacing $\Delta E$ $= E_L - E_H$. The magnitude of $T_c$ for a given nanocluster depends on these parameters and on the values of $\lambda_b$, $E_F$ , and $\tilde{\Omega}$, which are already known for each material. Remarkably, for perfectly realistic values of these parameters , a high value of $T_c$ can be obtained. Consider, for example, a cluster with the following parameter values:

$\Delta E = 65$ meV, $\tilde{\Omega} = 25$ meV, $m^* = m_e$, $k_F = 1.5 \times 10^8$ cm$^{-1}$, $\lambda_b = 0.4$,

radius R = 7.5 Å, and $G_H + G_L = 48$ (e.g., $l_H = 7$, $l_L = 4$).

For this set of values, one obtains $T_C \cong 10^2$ K(!). The large degeneracies of the highest occupied shell (HOS) and the lowest unoccupied shell (LUS) play an important role. Qualitatively, these degeneracies increase the effective electron-vibrational coupling $g_{eff}$ and, more specifically, the effective density of states. In principle, one can raise $T_c$ up to room temperature.

If we consider specifically a $Ga_{56}$ cluster (the Ga atom has three valence electrons, so that N=168), one can use the values $\tilde{\Omega} \approx 270$ K , $\lambda_b \approx 0.4$ , $m^* \approx 0.6 m_e$, and $k_F = 1.7 \cdot 10^8 cm^{-1}$. The calculation leads to $T_c \approx 145$ K(!) ,which greatly exceeds the bulk value ($T_c^b \approx 1.1$K).

It is important to stress that these high values of $T_c$ are caused by the electron-vibrational interaction.

A remaining question is how can one observe the appearance of pairing in an isolated cluster? Pairing leads to a strong temperature dependence of the excitation spectrum. Below $T_C$ and especially at low temperatures close to T=0 K, the excitation energy is strongly modified by the gap parameter and noticeably exceeds that in the region $T>T_c$. For example, the minimum absorption energy for $Gd_{83}$ clusters at $T>T_c$ corresponds to $\hbar\omega \approx 6$ meV, whereas for $T<<T_c$ its value is much larger: $\hbar\omega \approx 34$ meV. Such a large difference can be observed experimentally and is a manifestation of the superconducting state. It would be interesting to perform such experiments.

Recently Cao et al. (2008) used a specially developed technique (Breaux et. al., 2005) that allows one to measure the heat capacity of an *isolated* cluster. They observed jumps in heat capacity for selected Al clusters (e.g., for $Al_{35}^-$ ions) at T≈200K. The values of Tc as well as the amplitude of the jump and its width are in good agreement with the theory.

An anomalous diamagnetic moment can be also observed. In principle, a tunneling network of such nanoclusters can be built, and a macroscopic superconducting current could be observed.

## IX. Conclusion

In this Colloquium, we have described  as comprehensive as possible our view regarding the role of the electron-lattice (phonon) interaction in a number of novel superconducting systems , paying special attention to the cuprates.  We have indicated how this interaction can give rise to high temperature superconductivity, and we showed theoretically that even room temperature is possible within this framework. We have presented a variety of experimental observations that are consistent with this view. Furthermore, we have described a set of experiments for the cuprates that can provide an unambiguous answer to the question of the pairing boson .We hope that these experiments will be carried out in the near future.

Theoretically, the superconducting state can occur not only through the exchange by phonons, but also with the help of various bosons (e.g, of magnons). Only some experiments can dissociate between various channels and rule out those that do not provide any noticeable contribution. There has been sufficient experimental evidence for the importance of the electron-lattice (phonon) interaction. We think that the proposed experiments will provide additional crucial evidence for the concepts described above.

**Acknowledgments.**

We are grateful to R. Dynes and H. Morawitz for interesting discussions, and to L. Friedersdorf for help in the preparation of the manuscript. The research of VZK is supported by DARPA.

**Figure captions**

Fig.1. Structure of the Y-Ba-Cu-O (YBCO) compound.One can see

the apical, in–plane, and in-chain oxygen ions.

Fig.2. Function $\alpha^2(\Omega)F(\Omega)$ for Pb

Fig.3. Function $\alpha^2(\Omega)F(\Omega)$ for YBCO . From Dynes et al.,1992

Fig.4. I(V),dI/dV and $d^2I/dV^2$-characteristics for a BSCCO break

junction . From Aminov et al., 1994

Fig.5. Electronic terms (diabatic representation)

Fig.6. "Double-well" structure for the apical oxygen . From Haskel

et al.,1997

Fig.7. Plasmon spectrum for a layered electron gas

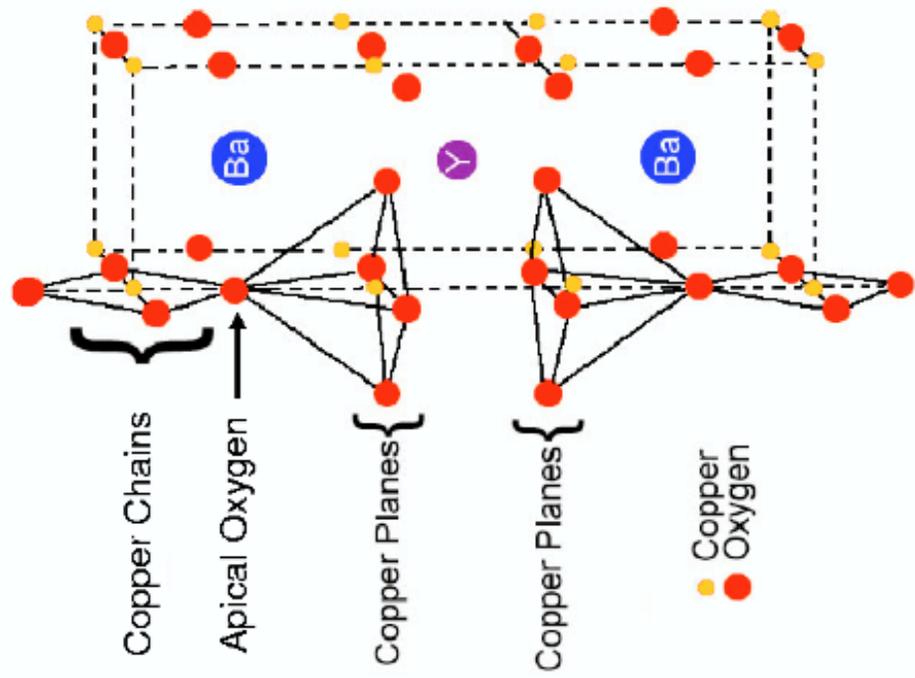

Figure 1.

Copper Chains

Apical Oxygen

Copper Planes

Copper Planes

Copper
Oxygen

Ba    Y    Ba

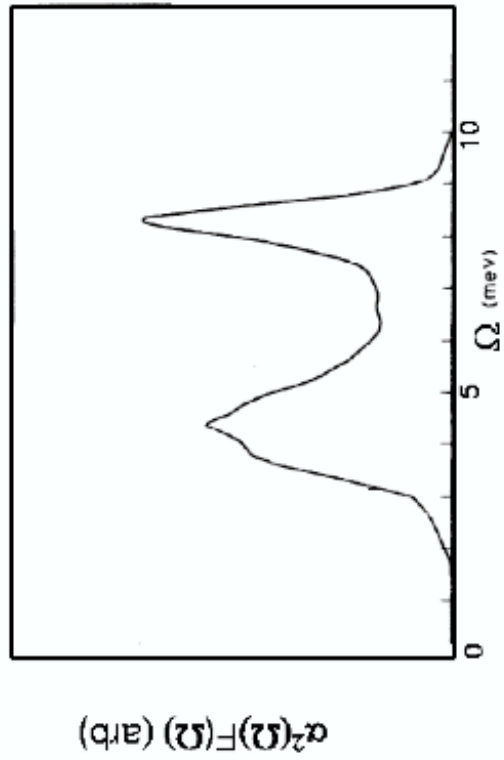

Figure 2

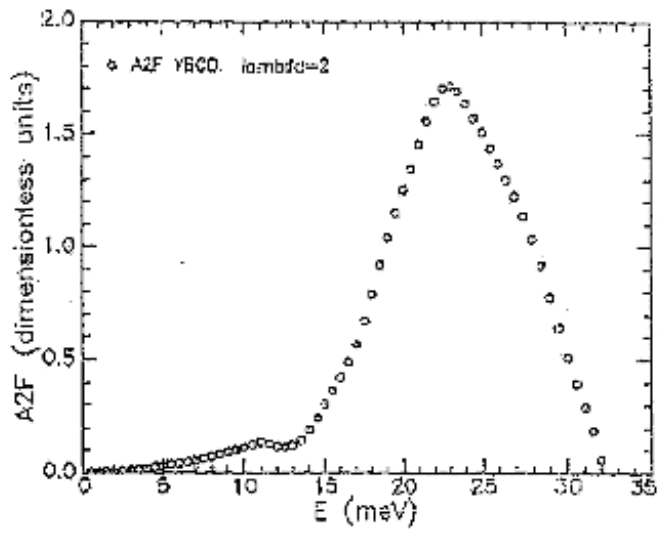

Figure 3

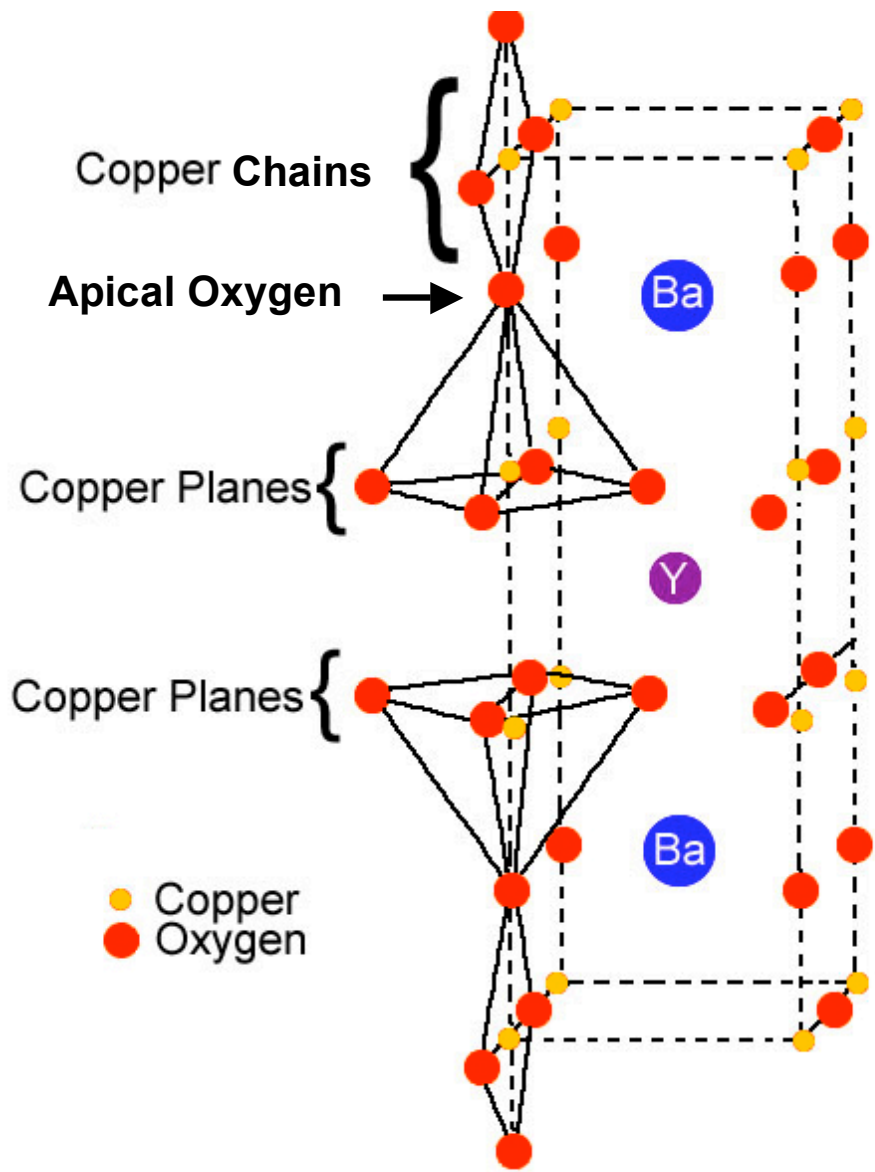

Copper **Chains** {

**Apical Oxygen** →

Ba

Copper Planes {

Y

Copper Planes {

Ba

● Copper
● Oxygen

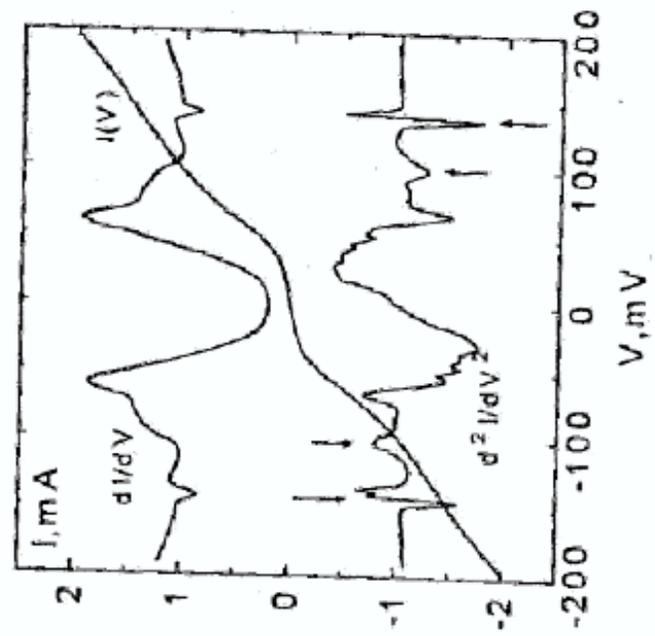

Figure 4

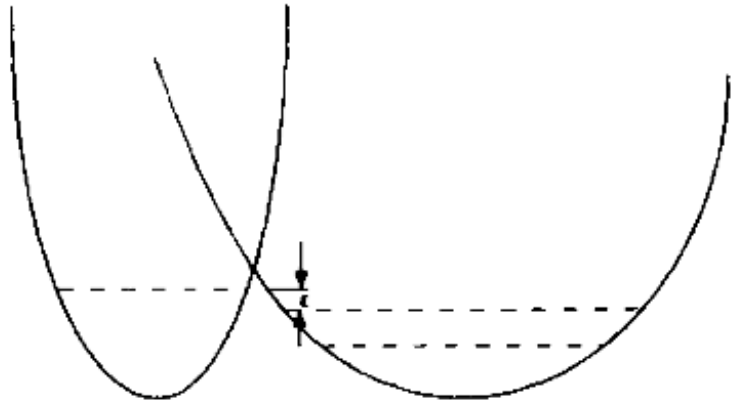

Crossing of Terms

Figure 5

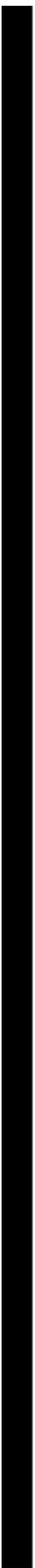
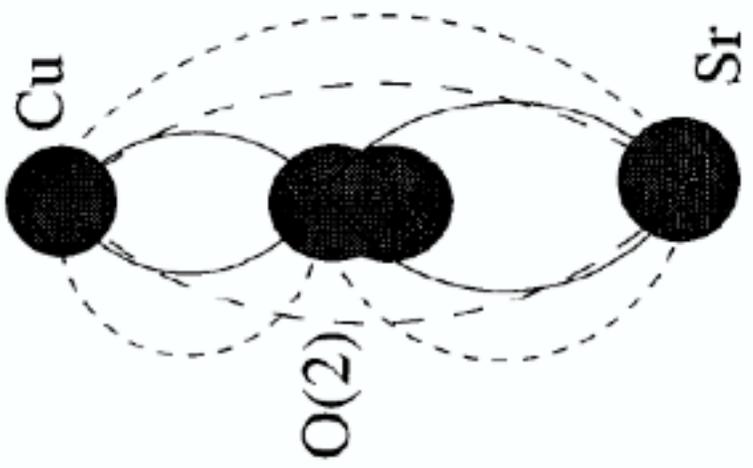

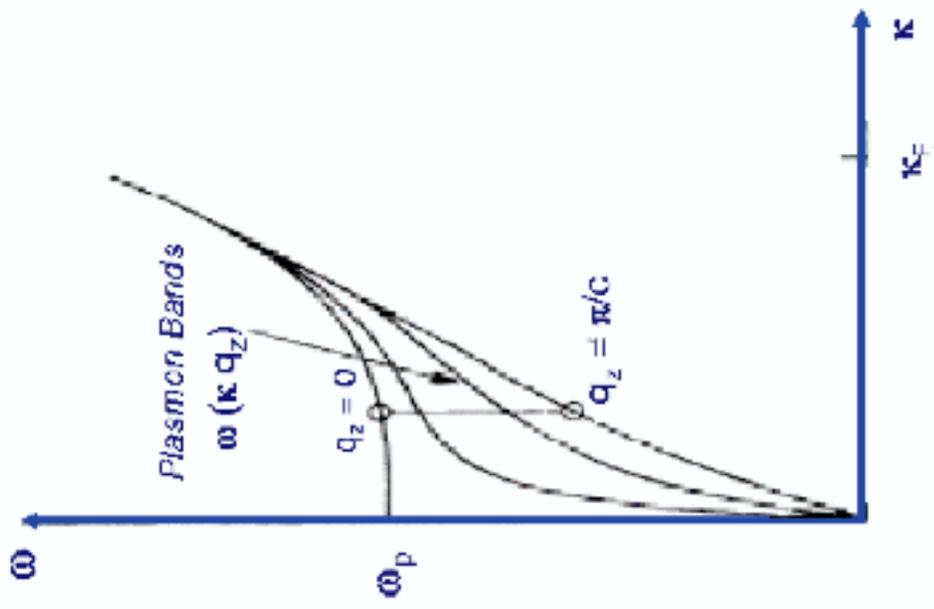
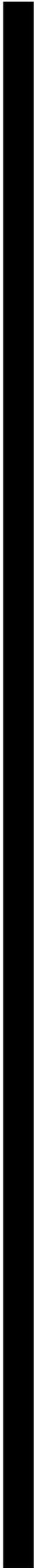